\documentclass[fleqn,usenatbib]{mnras}

\usepackage{newtxtext,newtxmath}
\usepackage[T1]{fontenc}
\usepackage{ae,aecompl}
\usepackage{hyperref}

\usepackage{graphicx}	% Including figure files
\usepackage{amsmath}	% Advanced maths commands

\usepackage{amssymb}	% Extra maths symbols
\usepackage{xcolor}
\usepackage{hyperref}
\usepackage{orcidlink}

\newcommand{\orcid}[1]{\href{https://orcid.org/#1}{\textcolor[HTML]{A6CE39}{\aiOrcid}}}

\def\la{\mathrel{\mathpalette\fun <}}
\def\ga{\mathrel{\mathpalette\fun >}}
\def\fun#1#2{\lower3.6pt\vbox{\baselineskip0pt\lineskip.9pt
        \ialign{$\mathsurround=0pt#1\hfill##\hfil$\crcr#2\crcr\sim\crcr}}}

\title[HOD modeling systematics]{Testing the framework of the halo occupation distribution with assembly bias modeling and empirical extensions}

\author[Z. Zhai et al.]{\parbox{\textwidth}{
Zhongxu Zhai,$^{1,2,3,4}$\thanks{E-mail: zhongxu.zhai@uwaterloo.ca}
Will J. Percival, $^{3,4,5}$
}\vspace*{4pt}\\
$^{1}$Department of Astronomy, School of Physics and Astronomy, Shanghai Jiao Tong University, Shanghai 200240, China \\
$^{2}$Shanghai Key Laboratory for Particle Physics and Cosmology, Shanghai 200240, China \\
$^{3}$Waterloo Center for Astrophysics, University of Waterloo, Waterloo, ON N2L 3G1, Canada \\
$^{4}$Department of Physics and Astronomy, University of Waterloo, Waterloo, ON N2L 3G1, Canada \\
$^{5}$ Perimeter Institute for Theoretical Physics, 31 Caroline St North, Waterloo, ON N2L 2Y5, Canada}

\date{Accepted XXX. Received YYY; in original form ZZZ}

\pubyear{2020}

\begin{document}
\label{firstpage}
\pagerange{\pageref{firstpage}--\pageref{lastpage}}
\maketitle

\begin{abstract}

We investigate theoretical systematics caused by the application of the halo occupation distribution (HOD) to the study of galaxy clustering at non-linear scales. To do this, we repeat recent cosmological analyses using extended HOD models based on both the Aemulus and Aemulus $\nu$ simulation suites, allowing for variations in the dark matter halo shape, incompleteness, baryonic effects and position bias of central galaxies. We fit to the galaxy correlation function including the projected correlation function, redshift space monopole and quadrupole, and consider how the changes in HOD affect the retrieval of cosmological information. These extensions can be understood as an evaluation of the impact of the secondary bias in the clustering analysis. In the application of BOSS galaxies, these changes do not have a significant impact on the measured linear growth rate. But, we do find weak to mild evidence for some of the effects modeled by the empirical parameterizations adopted. The modeling is able to make the HOD approach more complete in terms of cosmological constraints. We anticipate that the future and better data can provide tighter constraints on the new prescriptions of the HOD model. 

\end{abstract}


\section{Introduction}

Galaxy clustering measurements have wide application in cosmology and galaxy science. At large scales, the spatial distribution of galaxies reveals coherent patterns through baryon acoustic oscillations (BAO) and redshift space distortions (RSD). Observational data from cosmological surveys such as the Sloan Digital Sky Survey (SDSS-I/II, \citealt{SDSS_York, Abazajian_2009}), the Two degree Field Galaxy Redshift Survey (2dFGRS, \citealt{Colless_2001, Cole_2005}), WiggleZ (\citealt{Drinkwater_2010}), SDSS-BOSS (\citealt{Dawson_BOSS}), SDSS-eBOSS (\citealt{eBOSS_Dawson}), and the Dark Energy Spectroscopic Instrument (DESI, \citealt{DESI_2016, DESI_2024_BAO}) have provided accurate measurements of the expansion history and structure growth of the universe. When combined with perturbation theory, these measurements can constrain a large family of dark energy and modified gravity models (\citealt{Clifton_2012, Zhai_2017c, Abdalla_2022, Ferraro_2022}).

At small scales, the clustering amplitude is also sensitive to the physics underlying galaxy formation and evolution (see, e.g., \citealt{Zheng_2007, White_2007, Brown_2008, Wechsler_2018, Salucci_2019} and references therein). Due to the dual dependence on cosmology and galaxy formation it is challenging, but not impossible, to extract cosmological information on small scales. One way to do this is to use N-body simulations to provide an accurate description of the non-linear evolution of the dark matter field. Approximate statistical methods can then be used to place galaxies within dark matter halos, which allows sufficient sampling in the parameter space to marginalize over the model dependence. This combined approach can allow us to extract cosmological information at non-linear scales, but requires N-body simulations spanning the range of cosmologies to be tested. In order to interpolate between a small number of such simulations, the emulator method (\citealt{Heitmann_2009}) has been demonstrated to provide accurate measurements for the growth rate of structure using the recent observational data (see \citealt{Kwan_2015, Zhai_2019, Chapman_2021, Zhai_2023a, Yuan_2022, Kwan_2023} and references therein), as well as investigations into other aspects (\citealt{Heitmann_2010, Wibking_2017, Knabenhans_2019, Rogers_2019, McClintock_2018, Nishimichi_2019, Lange_2021, Storey-Fisher_2022, Chapman_2023}).

In the application of the emulator approach to the SDSS-BOSS galaxies, \cite{Zhai_2023a} report that the amplitude of matter fluctuation is lower than that from the cosmic microwave background (CMB) observed by the Planck satellite (\citealt{Planck_2020}), a finding similar to other studies using galaxies at lower redshift (\citealt{Leauthaud_2017, Wibking_2020, Lange_2021}). This may imply that the modeling to connect galaxies with dark matter halos is not complete, or novel physics is required to reconcile the various experiments. This emulator approach relies on the halo occupation distribution (HOD) model to describe how galaxies are distributed within halos. The method is flexible and can be easily extended with more parameters to describe physics beyond its most basic form where halo mass is the only dominant factor (\citealt{Hearin_2016}). Indeed, \cite{Zhai_2023a} considered the freedom that the velocity field of central or satellite galaxies can be different from the dark matter field (\citealt{Guo_2015}), and the effect of assembly bias, where the clustering properties of galaxies or dark matter halos are affected by the local environment in addition to halo mass. They also adopted a velocity scaling parameter to enable a more flexible modeling of the velocity field to retrieve the information due to redshift space distortion (\citealt{Reid_2014}). 

In addition to the standard two point correlation function other summary statistics, providing extra small-scale non-linear information, can be explored with a similar emulator approach. These include the marked correlation function (\citealt{Storey-Fisher_2022}), void-based statistics \citep{Fraser2024}, the density split method for galaxy clustering (\citealt{Paillas_2024}), the wavelet scattering transform algorithm (\citealt{Valogiannis_2022, Valogiannis_2024}), and the $k-$th nearest neighbor statistics (\citealt{Yuan_2024}). When analysing the two point functions and these statistics, marginalising over the HOD parameters to remove the effects of galaxy formation relies on the HOD being a complete description of the link between galaxies and mass in the Universe. This motivates us to consider the impact of more complications in the model of the galaxy halo connection. This will also allow us to investigate the correlation between the empirical models of galaxy bias and the cosmological parameters, and see if any of the extensions we consider can mitigate the tension between the large-scale structure (LSS) measurement and CMB observations. We consider the parameterization and modeling of both the central and satellite galaxies, including the dark matter halo shape and distribution of central galaxies within the halos. We also test the assumption of the completeness of the HOD model by incorporating an additional parameter to describe model incompleteness. Although these effects may not directly result from the assembly history of dark matter halos, they are inspired by considering how galaxy formation might be affected by halo properties. From this point of view, any impact on the clustering property that is not fully modeled by halo mass could be understood as a secondary bias or assembly bias.

We note that these effects have been studied in the literature for different purposes. For instance, within HOD models, the simplest assumption is that haloes are spherical. However, we know that even simple models of structure formation lead to non-spherical collapse (\citealt{Zeldovich_1970, Doroshkevich_1970}). Furthermore, simulations show that mergers of dark matter halos can further change the internal structures. Early studies found that the dark matter halo shape can be described by a triaxial ellipsoid model with explicit dependence on halo mass and redshift (\citealt{Jing_2002, Allgood_2006}). Using the Millennium simulation and a galaxy catalog generated by a semi-analytic model, \cite{Zu_2008} and \cite{vanDaalen_2012} reported that the halo shape can significantly affect the galaxy correlation function in both real and redshift space, especially at small scales. More recently, \cite{Mezini_2024} studied the Milky Way sized and cluster sized dark matter halos and found the anisotropic and aligned distribution of subhalos within their host halos. \cite{Durkalec_2024} investigated the halo asymmetry using both a HOD and subhalo abundance matching method. This effect may directly influence the retrieval of cosmological information using galaxy clustering measurements at scales where the impact is significant. To explicitly quantify the potential effect, we incorporate this halo shape into our previous emulator-based modeling to explore the possible bias on cosmological parameter inference. In addition, we consider effects such as the spatial distribution of central galaxies, baryonic effects and incompleteness of the HOD model in this work. We also investigate the impact of neutrino mass using the new released Aemulus $\nu$ simulation (\citealt{DeRose_2023}). 

This paper is organized as follows. In Section 2, we give an overview of our modeling of galaxy clustering at small scales within the emulator approach. In Section 3, we describe how we extend the basic model with additional prescriptions. In Section 4, we discuss our results and conclude with a summary of the main findings.

\section{Modeling}

In this section we first introduce our analysis pipeline based on the HOD framework, including the summary statistics, model parameters and likelihood analysis. Then we focus on a new addition to the method, including the AP effect. 

\subsection{Basic framework}

The basic framework for the analysis follows from our previous work modeling galaxy clustering at small scales \citep{Zhai_2019, Zhai_2023a, Zhai_2023b}, and we only summarize it briefly in this section. We fit models to galaxy clustering measurements based on the two-point correlation function (2PCF) $\xi(r)$ which measures the excess probability of finding two galaxies separated by a distance $r$. In order to include information that does not depend on modeling the RSD effect, we fit the projected correlation function (\citealt{Davis_1983})
\begin{equation}\label{eq:wp}
    w_{p}(r_{p}) = 2\int_{0}^{\infty}d\pi \xi(r_{p}, \pi),
\end{equation}
where $r_{p}$ and $\pi$ denote the separations of the galaxy pair perpendicular and parallel to the line-of-sight respectively. We combine this with the clustering measurement in redshift space, decomposing the correlation function to compute the multipoles
\begin{equation}\label{eq:xi}
    \xi_{\ell} = \frac{2\ell+1}{2}\int_{-1}^{1}L_{\ell}(\mu) \xi(s, \mu) d\mu,
\end{equation}
where $L_{\ell}$ is the Legendre polynomial of order $\ell$, $s$ is the magnitude of the separation and $\mu$ is cosine of the angle between $r$ and the line-of-sight. We only consider the first two orders of the multipole $\xi_{0}$ and $\xi_{2}$. The measurements of these statistics are performed with 10 logarithmic bins for $r_{p}$ or $s$ from 0.1 to 60.2 $h^{-1}$Mpc. 

We model galaxy clustering using the HOD framework, based on the model of \cite{Zheng_2005}. The basic component is the mean number of central and satellite galaxies in each halo
\begin{eqnarray}\label{N_cen}
N_{\text{cen}}(M) &=& \frac{1}{2} \left[1+\text{erf} \left(\frac{\log_{10}{M}-\log_{10}{M_{\text{min}}}}{\sigma_{\log{M}}}\right)\right], \\
N_{\text{sat}}(M) &=& \left(\frac{M}{M_{\text{sat}}}\right)^{\alpha}\exp{\left(-\frac{M_{\text{cut}}}{M}\right)} N_{\text{cen}}(M),
\end{eqnarray}
where $M_{\text{min}}$, $\sigma_{\log{M}}$, $\alpha$, $M_{\text{sat}}$ and $M_{\text{cut}}$ are free parameters. This model has been used extensively for clustering analysis of massive galaxies, e.g. \cite{CMASS_Martin, Parejko_LOWZ, Zhai_2017}. To complete the model, we also consider parameters for additional effects, including the concentration parameter for the satellite distribution compared with the NFW profile (\citealt{NFW_1996}), velocity bias parameters for central and satellite galaxies (\citealt{Guo_2015}), incompleteness parameter for central occupancy at the massive end (\citealt{leauthaud_etal:16, Lange_2021}), scaling parameter of the velocity field of dark matter halos (\citealt{Reid_2014}) and environment-based assembly bias parameters (\citealt{Mcewen_2018, Han_2019, Salcedo_2018, Yuan_2020, Xu_2020}). We refer the readers to Table~3 of \cite{Zhai_2023a} for detailed definition and symbols of the model parameters used throughout the analysis. 

We use the above HOD model to populate dark matter halos from numerical simulations to generate mock galaxy catalogs. In this work, we utilize the Aemulus simulation suite (\citealt{DeRose_2018}) and the new Aemulus $\nu$ suite (\citealt{DeRose_2023}) with massive neutrinos. The Aemulus suite comprises 40 simulations with different cosmological parameters to construct the emulator, and another independent 35 simulations to evaluate the emulator accuracy. 

The Aemulus suite is based on dark matter only simulations without massive neutrinos. Modeling the massive neutrino component in numerical simulations for cosmological applications has been an active topic for years, and a variety of methods have been developed in the literature. Depending on the mass range and the required accuracy, both linear and non-linear treatments of the neutrino component have been applied to different purposes, see e.g., \cite{Brandbyge_2009, Viel_2010, Alihamoud_2013, Upadhye_2016, Banerjee_2016, Banerjee_2018}, and references therein. Research on large-scale structure shows that the finite mass of neutrinos leads to a large velocity dispersion and suppresses the clustering amplitude of dark matter halos (\citealt{Lesgourgues_2006, Saito_2008}). Given the measured precision of the growth rate using galaxy clustering at small scales from our previous work, it is worthwhile to quantitatively investigate the impact of neutrino mass. For this purpose, we utilize the latest Aemulus $\nu$ simulation suite \cite{DeRose_2023} in part of this work, which extends and updates the previous Aemulus suite while maintaining the same simulation volume and mass resolution. For more details on this simulation, we refer interested readers to \cite{DeRose_2023}.

In order to provide a reference model for comparison when we investigate the model extensions using simulations with massive neutrinos, we rebuild the emulator and rerun the analysis performed in \citet{Zhai_2023a}. In terms of the measurement of linear growth rate, we find that the massive neutrinos do not have a significant impact as we show later. We will present a thorough analysis for the constraint on the neutrino mass in a separate paper (\citealt{Gao_2024}), and we focus on the extension of our model of galaxy halo connection in this work.

The Aemulus $\nu$ suite has two tiers. Tier 1 includes 100 simulations covering a wider range of cosmological parameter space, while tier 2 comprises 50 simulations with similar coverage to Aemulus. To ensure emulator accuracy with dense samplings, we only use tier 2 for massive neutrino-related modeling in this work. Both the Aemulus and Aemulus $\nu$ simulations have the same box size of $1.05h^{-1}$ Gpc and $1400^3$ dark matter particles, while the Aemulus $\nu$ boxes also have $1400^3$ neutrino particles. The mass resolution slightly depends on the cosmological parameters but is suitable for massive galaxies like the BOSS sample. Both simulation suites are run with a spatially flat $w$CDM cosmology, and the model parameters with their ranges can be found in \cite{DeRose_2018, DeRose_2023} or Table 3 of \cite{Zhai_2023a}. The evaluation of each model extension is performed individually with a reference model. We provide more details in the corresponding sections and summarize in Table~\ref{tab:param}.

We use the Latin-hypercube algorithm (\citealt{Heitmann_2009}) to sample the HOD model parameter space and generate the training sample. The measurement of the two-point correlation function (2PCF) (Eq. \ref{eq:wp} and \ref{eq:xi}) from these mocks yields the final training set to construct the emulator, and the detailed methodology follows our earlier work (\citealt{Zhai_2019}).

In order to constrain cosmological parameters, we perform a Bayesian analysis, with a Gaussian likelihood
\begin{equation}
\ln{\mathcal{L}} = -\frac{1}{2}(\xi_{\text{emu}}-\xi_{\text{obs}})C^{-1}(\xi_{\text{emu}}-\xi_{\text{obs}})\,.
\end{equation}
Note that we include emulator inaccuracies in the analysis, using the test simulations to evaluate the emulator accuracy. We include the scatter as an additive contribution to the covariance matrix. To obtain the posterior of the model parameters, we use the nested sampling algorithm (\citealt{Skilling_2004}) provided by the MultiNest (\citealt{Feroz_2009, Buchner_2014}) package. The tests run until the convergence criteria are reached, i.e., the Bayes Evidence estimate is stable and the posterior can be produced as a byproduct.

\subsection{Alcock-Paczynski Effect}\label{sec:AP}

\begin{figure}
\begin{center}
\includegraphics[width=9cm]{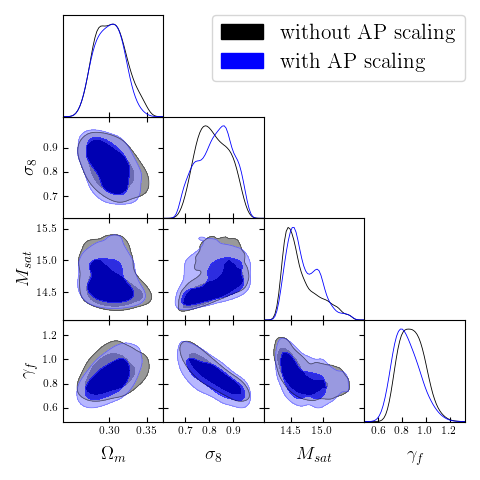}
\caption{Constraint on a subset of the model parameters with the AP effect corrected using BOSS galaxies. For reference, the original result without AP stretching of the models is also shown. The result does not reveal systematic offset due to the addition of the AP effect, and thus the measurement of growth rate is not significantly affected.}
\label{fig:AP_contour}
\end{center}
\end{figure}

For the measurement of the galaxy correlation function, we must assume a fiducial model to convert redshift into distance, which means the measured correlation function depends on this choice. To mimic this within the model to be fitted to the data, we need to stretch the simulations we use along and across the line-of-sight by the Alcock-Paczynski parameters (AP parameters, \citealt{Alcock_1979}). 

In the eBOSS analysis (\citealt{Bautista_2021, Chapman_2021}), this effect was incorporated by scaling the distance parameter when making predictions of the two-point correlation function, but that ignores the effect on halo finding. In our work, instead, we include the effect during the construction of the galaxy mocks, following \cite{Lange_2023}, where the simulation coordinates are adjusted based on the assumed fiducial cosmology. 

Before we apply the AP correction to the model, we first examine the effect on the correlation function measurement. In order to do so, we define two parameters $q_{\parallel}$ and $q_{\perp}$ and use them to scale the axes for LoS and perpendicular directions respectively, to mimic the AP effect. 
These parameter are not the same as the $\alpha$s commonly used to compress Baryon Acoustic Oscillation (BAO) measurements, as they do not include the sound horizon at the drag epoch: We are interested in the shift in the full clustering signal caused by the mismatch of true and fiducial cosmology, not the shift of the BAO position from changes in both the geometric stretching and the sound horizon. In order to maintain the galaxy number density, we regenerate the HOD mocks using the Aemulus boxes with the number density corrected by the factor $q_{\parallel}q_{\perp}^{2}$ such that the resultant mock after AP parameter correction matches the observed galaxy number density. 

Given this modeling, we can test the impact of the AP effect on the cosmological constraints. Since the cosmological parameters are allowed to vary, we can explicitly compute the strength of the AP correction for each corresponding model being tested. Therefore $q_{\parallel}$ and $q_{\perp}$ are not free parameters. To implement the AP effect, we construct the emulators for the correlation function with $q_{\parallel}$ and $q{\perp}$ as part of the extended parameter set, such that we do not have to define a fiducial cosmology. Then, in the likelihood analysis when the fiducial cosmology is known, we compute the AP parameters of each model as derived parameters and use these values. In Figure~\ref{fig:AP_contour} we present the constraint on a subset of parameters in the cosmology and HOD parameter set. For reference, we also show the result when the AP scaling is not applied. We can see that these AP parameters do not have a significant impact on the cosmological measurements. We also compute the linear growth rate and find that the offset is no higher than a few tenth of $\sigma$. Therefore we conclude that ignoring AP stretching in the analysis using small scale clustering do not have a significant impact. This is consistent with the earlier eBOSS result but using a different approach (\citealt{Chapman_2021}).

\section{Extensions}

\begin{table*}
\centering
\begin{tabular}{llllcr}
\hline
& Parameter  & Meaning & Range  & Simulation  & Section\\
\hline
 & $q_{\parallel}$, $q_{\perp}$  & AP parameters    & derived & Aemulus &  \ref{sec:AP} \\ 
 & $e_{1}, e_{2}$  & shape parameter for dark matter halo    & [0, 1] & Aemulus &  \ref{sec:shape} \\
 & $L_{frac}$   & HOD incompleteness parameter   & [0, 0.2] & Aemulus $\nu$ &  \ref{sec:incompleteness} \\ 
 & $B_{f}$   & scaling parameter for baryonic effect  & [0.5, 1.5] & Aemulus $\nu$ &  \ref{sec:baryon} \\ 
 & $p_{b}$   & position bias for central galaxies  & [0, 0.3] & Aemulus $\nu$ &  \ref{sec:central} \\ 
\hline
\end{tabular}
\caption{Systematics and Model extensions investigated in this work. We list the symbols for the parameters, physical meaning, range in the modeling, simulation suite used, and the corresponding section in this paper.}
\label{tab:param}
\end{table*}

In this section, we introduce the extensions considered for the HOD-based model. In Table~\ref{tab:param}, we summarize the key parameters and their corresponding information. 

\subsection{Dark matter halo shape}\label{sec:shape}

The first extension we consider is the effect of dark matter halo shape on clustering measurements using high-resolution N-body simulations. We then introduce a parameterization to extend the parameter space of the HOD-based model, and finally present the impact on cosmological parameters.

\subsubsection{Impact of halo shape on correlation function}

\begin{figure*}
\begin{center}
\includegraphics[width=18cm]{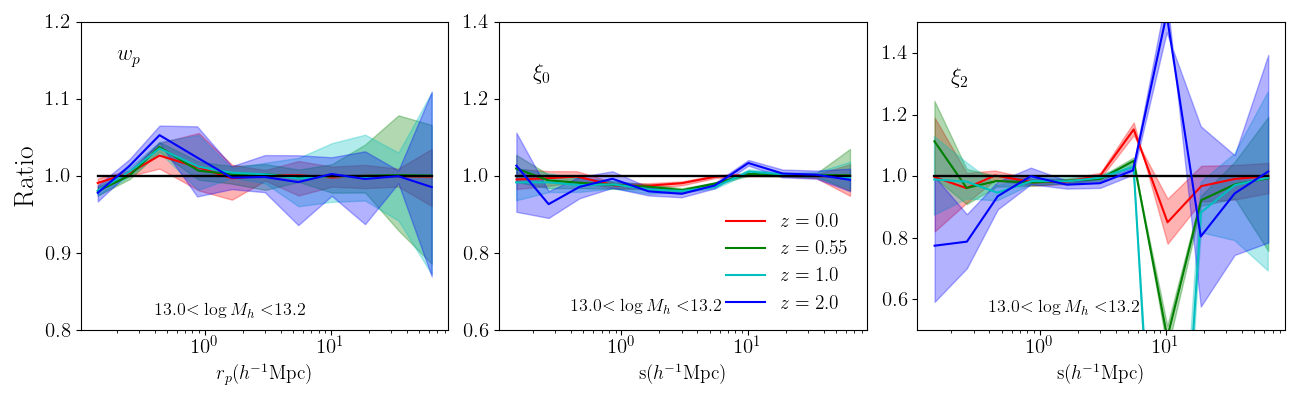}
\includegraphics[width=18cm]{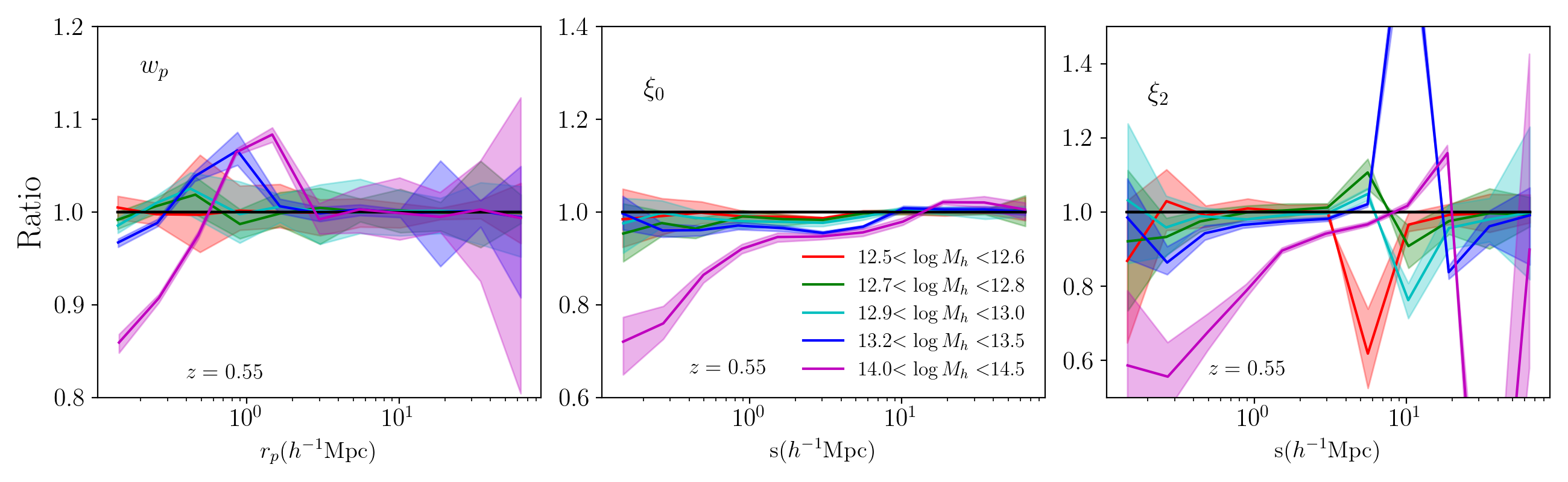}
\caption{Impact of changes in the dark matter halo shape assumed on the correlation function in redshift space: $w_{p}$ (left), RSD monopole $\xi_{0}$ (middle) and quadrupole $\xi_{2}$ (right), using the halo and subhalo catalog from the MDPL2 simulation. The result shows the ratio of the measurements between sphericalized and the original distribution. The top panel shows a sample with halo mass $13.0<\log{M[h^{-1}M_{\odot}]}<13.2$ at different redshifts, while the bottom panel shows samples at $z=0.55$ with different halo masses. For each realization, we repeat the sphericalization procedure multiple times with different random seeds and measure the 2PCF. The solid line denotes the mean and the shaded area displays the 68\% uncertainty from these measurements. }
\label{fig:cf_shape}
\end{center}
\end{figure*}

\begin{figure}
\begin{center}
\includegraphics[width=9cm]{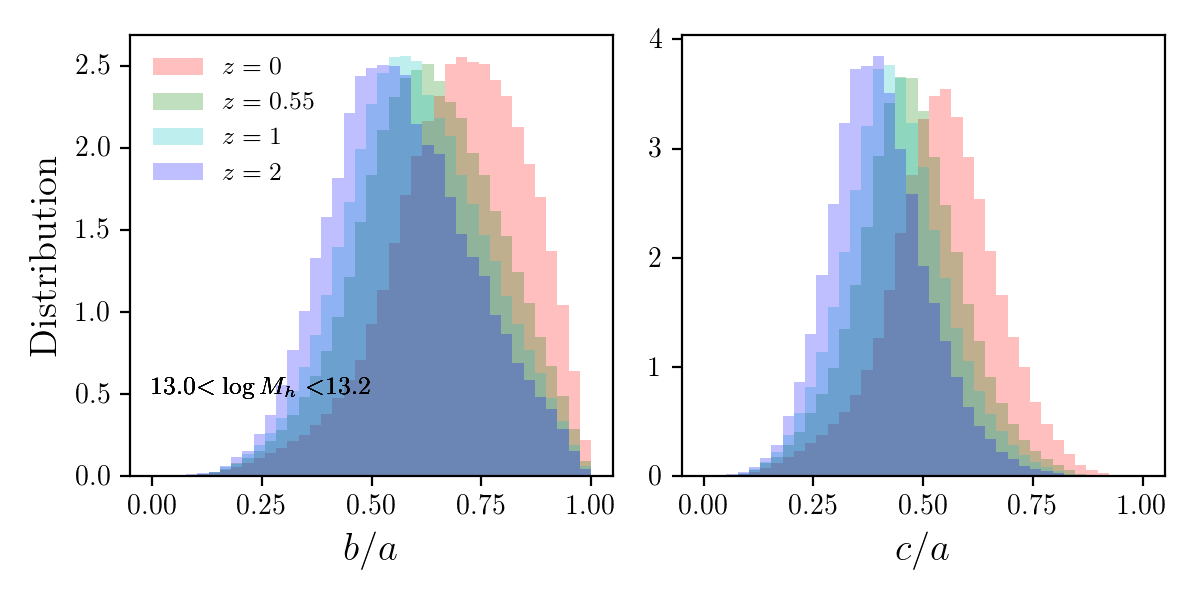}
\includegraphics[width=9cm]{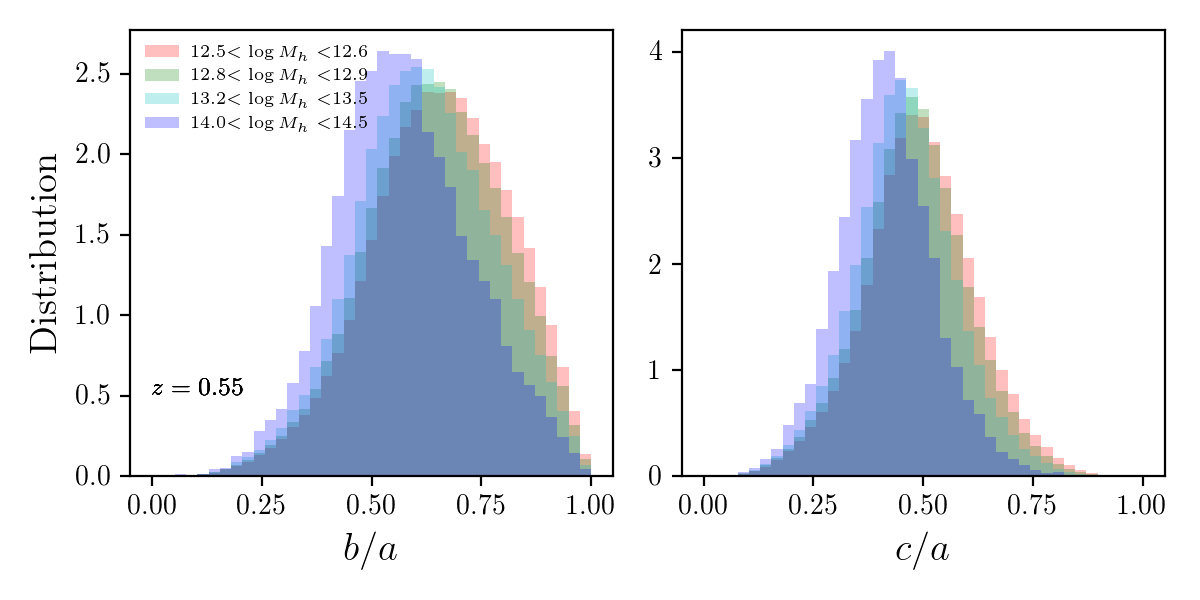}
\caption{The distribution of the axis ratios for dark matter halos identified in the MDPL2 simulation. The top row shows the redshift dependency of a sample selected by mass, while the bottom row shows the mass dependency at z=0.55.}
\label{fig:ellipticity}
\end{center}
\end{figure}

We adopt a simple approach using subhalos instead of galaxies such that the halo shape is represented by the spatial distribution of the subhalos within the host halo. The halo catalog we use is from the MDPL2 simulation (\citealt{Prada_2012, Klypin_2016})\footnote{\url{https://www.cosmosim.org/metadata/mdpl2/}} with cosmological parameters $\{\Omega_{m}, \Omega_{b}, h, n_{s}, \sigma_{8}\} = \{0.307, 0.048, 0.677, 0.96, 0.8228\}$. The simulation has a box size of $1h^{-1}$ Gpc with $3840^3$ dark matter particles, and the mass resolution is $1.51 \times 10^9h^{-1}M_{\odot}$. The dark matter halos are identified using the Rockstar algorithm (\citealt{Behroozi_2013b}). In order to avoid artifacts due to unresolved (sub-)structures, we restrict the analysis in this section to only (sub)halos above $10^{12}h^{-1}M_{\odot}$, which corresponds to at least one hundred dark matter particles as the lower limit.

We test the effect of halo shape using the method from \cite{Zu_2008} and \cite{vanDaalen_2012}. To form a baseline, we randomly rotate the position of each individual subhalo around the center of the host halo, i.e., we sphericalize the spatial distribution of subhalos within the host halo. We also apply the same rotation vector to the relative velocity of the subhalos with respect to the center of the host halo. This implicitly keeps the correlation between the host halo and subhalos, such as the infall pattern. In Figure~\ref{fig:cf_shape}, we present the ratio of the 2PCF between the sphericalized and the original distribution measured in redshift space for $w_{p}$, $\xi_{0}$, and $\xi_{2}$ respectively. The top row shows the example when we only use halos with a mass in the range $13.0 < \log_{M} < 13.2$, but for different redshifts, and the bottom row picks the $z=0.55$ snapshot with different ranges of halo mass. For each realization, we repeat the sphericalization procedure twenty times and measure the 2PCF with three axes as the line of sight. The solid line denotes the mean, and the shaded area displays the 68\% uncertainty from these 60 measurements.

From the results, we can see that the dark matter halo shape can impact the clustering measurement mainly at small scales, consistent with previous studies and our expectations, since the sphericalization process only redistributes the subhalos (satellite galaxies). We see some change in $\xi_{2}$ at $\sim10~h^{-1}$ Mpc, but it appears relatively stronger than it is, due to the statistics crossing zero. From the top row, the results show some dependence on redshift, i.e., halos of the same mass can experience more effects due to shape in terms of both the amplitude and shape of the correlation function at higher redshift. However, this dependence is not as strong as the halo mass dependence as shown in the bottom row. More massive halos have a stronger effect due to shape. The effect on the most massive halos can become significant at a few $h^{-1}$ Mpc, simply due to the increase in the halo radius.

This dependency can be partially explained by the distribution of the shape parameters of dark matter halos. Figure~\ref{fig:ellipticity} shows the ellipticity parameter of the Rockstar halos defined using the mass distribution tensor for particles within the halo radius (\citealt{Zemp_2011, Behroozi_2013b})
\begin{equation} \label{eq:Mij}
    M_{ij} = \frac{1}{N}\sum_{N}x_{i}x_{j},
\end{equation}
where $x_{i}$ denotes the $i$-component of the position vector. We can obtain the sorted eigenvalues of this matrix to get the axes of the principal ellipsoid ($a>b>c$). This matrix indeed simplifies the shape information of the halo into a few parameters defining an ellipsoid. We show the distribution of this result using the MDPL2 simulation in the figure for various masses and redshifts. For both axis ratios ($b/a$ and $c/a$), we can see a similar dependency with slightly different amplitudes of change. The result is consistent with early studies \cite{Jing_2002, Kasun_2005, Allgood_2006, Bett_2007, Bonamigo_2015, VegaFerrero_2017}, i.e., the axis ratio is larger (closer to sphericity) for less massive halos and lower redshifts. This dependency can be further investigated within the framework of hierarchical growth of structure formation, but the scatter of the distribution may imply complicated couplings of various halo properties, such as the halo formation history, mass assembly history, baryonic physics, local environments, and so on. We refer the readers to e.g., \cite{Smith_2006, JeesonDaniel_2011, Bryan_2013, Chua_2019, Chen_2020, Lau_2021, Cataldi_2023} and references therein for some recent and further studies.

\subsubsection{Modeling of halo shape in HOD framework}

When using the HOD model to create mock catalogues, the simplest approaches to implement halo shape are to either place satellites such that they are spherically distributed or to place satellites that follow dark matter particles. Both methods have been used extensively in the past, but we should note that this binary choice may not be flexible enough given the required accuracy of LSS measurements from future surveys with larger volume and higher number density. In particular, the baryonic physics of galaxy formation may impact the evolution of satellite galaxies within their host halos such that the distribution does not match that of the dark matter. There can be plenty of effects or phenomena caused by the baryonic processes, and the lack of a well-established model of galaxy formation can make an exhaustive exploration impossible. Therefore, in this work, we use the halo shape as an example and implement an empirical model given the emulator approach we have developed. i.e. we can parameterise the shape and test if the data shows any evidence for non-standard behaviour.

We use the distribution of satellite galaxies to represent the dark matter halo shape. To implement the model, we first use the HOD formalism to generate galaxy mocks assuming the satellites are spherically distributed as in our previous work. Then for each host halo, we scale the axis ratios ($b/a$ and $c/a$) determined through the ellipsoidal model (Eq. \ref{eq:Mij}) by introducing two new parameters:
\begin{equation}\label{eq:shape1}
    i_{1} = (b/a)^{e_{1}}, \quad i_{2} = (c/a)^{e_{2}}.
\end{equation}
With the re-scaled axis ratio, we scale the coordinates of the satellites in the host halo frame through
\begin{eqnarray}\label{eq:shape2}
    x' & = & x / (i_{1}i_{2})^{1/3}, \nonumber \\
    y' & = & y i_{1} / (i_{1}i_{2})^{1/3}, \\
    z' & = & z i_{2} / (i_{1}i_{2})^{1/3}. \nonumber
\end{eqnarray}
Note that we also apply this scaling method to the velocity components of the satellites in the host halo frame. The final step is to perform a single rotation to all of the satellites belonging to the host halo, and we repeat the process for all the host halos. This removes the correlation between halo orientation and environment, and allows us to consider the effect of the radial distribution only.  In the parameterization we choose, haloes can return to the spherical shape when $e_{1} = e_{2} = 0$, and to an ellipsoidal shape when $e_{1} = e_{2} = 1$. Other values of $e_{1}$ and $e_{2}$ represent the deviation of the halo shape from spherical or ellipsoidal. We note that this parameterization is not unique, as one can also adopt models for triaxiality to describe ellipticity and prolaticity as used in e.g., \cite{Franx_1991, Jing_2002, Allgood_2006, Lau_2021}.

\begin{figure}
\begin{center}
\includegraphics[width=9cm]{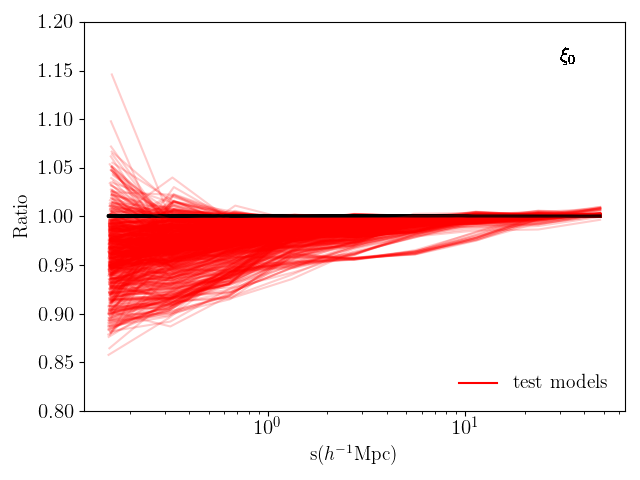}
\caption{Effect of our parameterization of dark matter halo shape on $\xi_{0}$ in redshift space, expressed as the ratio between shape modulated and the original spherical distribution. It displays results from hundreds of HOD models populating the Aemulus test simulations. }
\label{fig:shape_xi0}
\end{center}
\end{figure}

\begin{figure}
\begin{center}
\includegraphics[width=8cm]{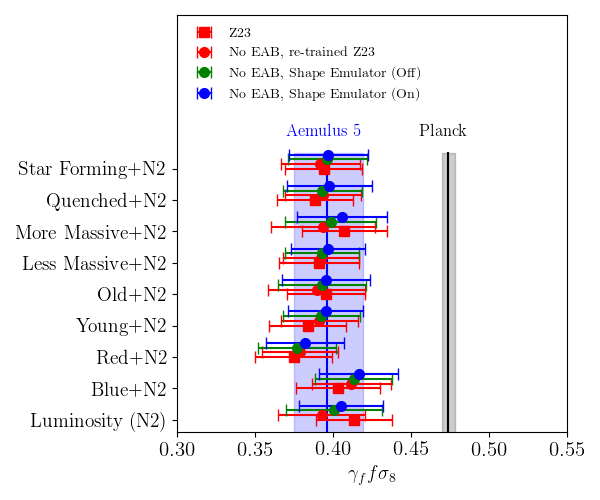}
\caption{Measurement of $f\sigma_{8}$ using different emulator models. For reference, the original result from \protect\cite{Zhai_2023b} is shown as red squares, the result using the same model but with assembly bias turned off is shown as red dots. The results from the new emulator with varying halo shape modelling are shown as blue (shape parameter added) and green (spherical distribution) respectively.}
\label{fig:shape_growth}
\end{center}
\end{figure}

To see the impact of this model, we apply this parameterization to the test samples of the Aemulus suite with large volumes and present the results of $\xi_{0}$ for 100 independent HOD models with various cosmological simulations in Figure~\ref{fig:shape_xi0}. The result is represented as the ratio of $\xi_{0}$ with the new satellite distribution divided by the measurement of the original HOD mock. We can see that these additional degrees of freedom can make the clustering amplitude lower than assuming a spherical distribution of satellites. The impact is clearly scale-dependent and more significant in the one-halo regime as expected. The overall result shows that the ellipsoidal halos are less clustered than spherical halos. This result is related to the work of \cite{Faltenbacher_2010}, which showed that near-spherical halos are more clustered than non-spherical ones. Their study was in the framework of assembly bias rather than empirical parameterization: they simply split the galaxy population and comparing their clustering amplitude. This result should not be confused with the process of sphericalization, as shown in Figure~\ref{fig:cf_shape}. Sphericalization can reduce the clustering amplitude for the same statistics ($\xi_{0}$). This is likely caused by the rotation operation increasing the separation of satellites within the dark matter halos on average \citep{vanDaalen_2012}. Our model works in a different way with the parameterization in Eq.~(\ref{eq:shape1}) and~(\ref{eq:shape2}) preserving the volume of the spheroid and ellipsoid, and thus the average separation of the galaxy pairs of the one-halo terms. The resultant behavior of the clustering statistics becomes closer to the physical exploration in \cite{Faltenbacher_2010}. We note that compared with those extensive discussions of the dependency on halo mass, concentration, and halo spin, our parameterization only serves as a simplified approximation.

We next apply this model to the training set of the HOD mocks and rebuild the emulator within the extended model parameter space. With the new parameters, the overall accuracy of the emulator remains comparable to the original one without significant degradation. We then apply the model to the BOSS measurement from \cite{Zhai_2023b} (hereafter Z23) at $z=0.55$ for subsamples split by different galaxy properties. To isolate the effect of the halo shape parameters, we do not implement the environment-based assembly bias parameters as in \cite{Zhai_2023a} and \cite{Zhai_2023b}, since we find that this effect is not significant in the BOSS galaxies. For a fair comparison, we first turn off the assembly bias parameter by setting $f_{\text{env}}=0$ in the original model (see e.g., \cite{Zhai_2023a} for more details of this model) and rerun the cosmological constraint. The result in Figure~\ref{fig:shape_growth} shows that ignoring this effect doesn't affect the measurement of $f\sigma_{8}$ (red square and red dot), as we already point out. We then apply the new emulator model with halo shape parameters to the same BOSS measurement (blue dot). The $f\sigma_{8}$ value is consistent with the previous one, indicating that the new degree of freedom does not prefer a higher $f\sigma_{8}$ measurement to resolve the discrepancy with Planck. It's likely that the shape model in the one-halo regime is not correlated strongly with cosmological information, or it's been obscured by significant shot noise. The former can be understood through the baryonic processes within the dark matter halos, while the latter can be further tested with future data such as DESI with higher galaxy number density (\citealt{DESI_2016}). We also perform a sanity check for the emulator construction. The original emulator with assembly bias parameters turned off is equivalent to the new emulator but with shape parameters turned off. The latter can be simply achieved by setting $e_{1}=e_{2}=0$, and we find that the result (green dot) is consistent with the previous ones, showing that the tiny variation is only due to the interpolation accuracy of the Gaussian Process.

\subsection{Incompleteness}\label{sec:incompleteness}

As mentioned in the previous sections, we use the Aemulus $\nu$ suite in the following analyses. The first extension we consider is the (in)completeness of the galaxy sample in terms of clustering measurement, which has been previously discussed in the literature such as \citet{leauthaud_etal:16, Tinker_2017a} and \citet{Zhai_2017}. To minimize the bias in the analysis of Aemulus V, we adopt corrections from both the sample and modeling sides. For the former, we restrict the sample to a narrow redshift range and only select galaxies at the brighter end. For the latter, we introduce the parameter $f_{\text{max}}$ to model the incompleteness of central galaxies at the massive end. Indeed our analysis in \cite{Zhai_2023a} has shown a correlation between the parameter $f_{\text{max}}$ and growth rate-related parameters. Enforcing this parameter to unity can lead to a lower estimate of $f\sigma_{8}$ with a significance of a few tens of $\sigma$, similar to the findings in \cite{Lange_2021}.

In \cite{Zhai_2023b}, we tried multiple selections on the galaxy sample in addition to their brightness. We note that value-added parameters such as star formation rate and stellar mass can lead to different levels of completeness in individual samples. This effect may not be fully corrected within the current HOD framework. Given the high accuracy of cosmological measurements, it is useful to revisit the effect. In this section, we focus on the modeling side and explore the systematics that may not be corrected by the current HOD model. On the other hand, there are artifacts in the numerical simulations where galaxies can be lost during the evolution of dark matter halos. These effects can introduce an extra source of scatter in the modeling of HOD. Therefore, we extend our model by introducing additional degrees of freedom and evaluate the impact on cosmological measurements. We define a new parameter $L_{frac}$ in this approach and discuss the modeling and results in the following four cases.

\begin{figure*}
\begin{center}
\includegraphics[width=8cm]{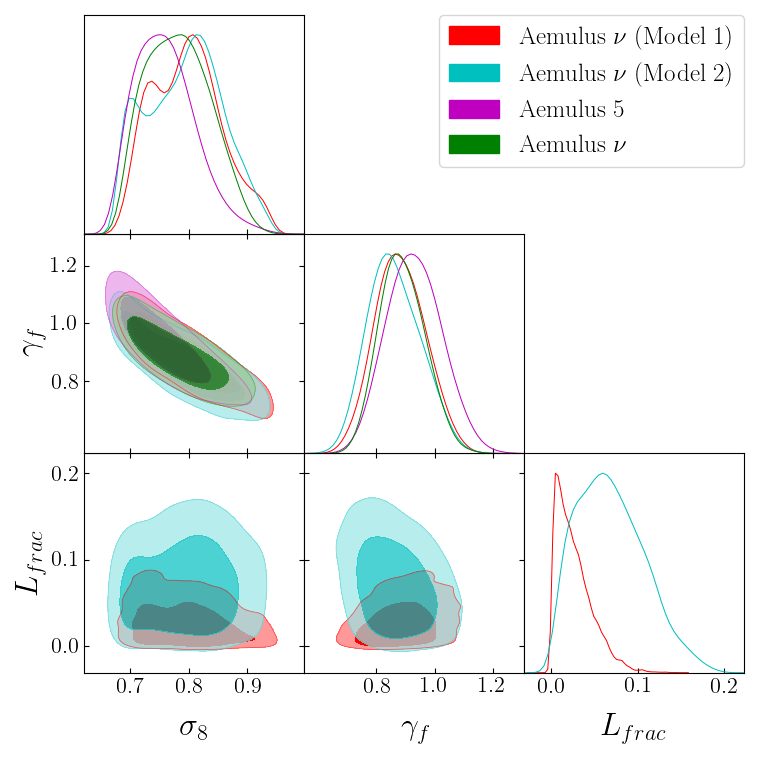}
\includegraphics[width=8cm]{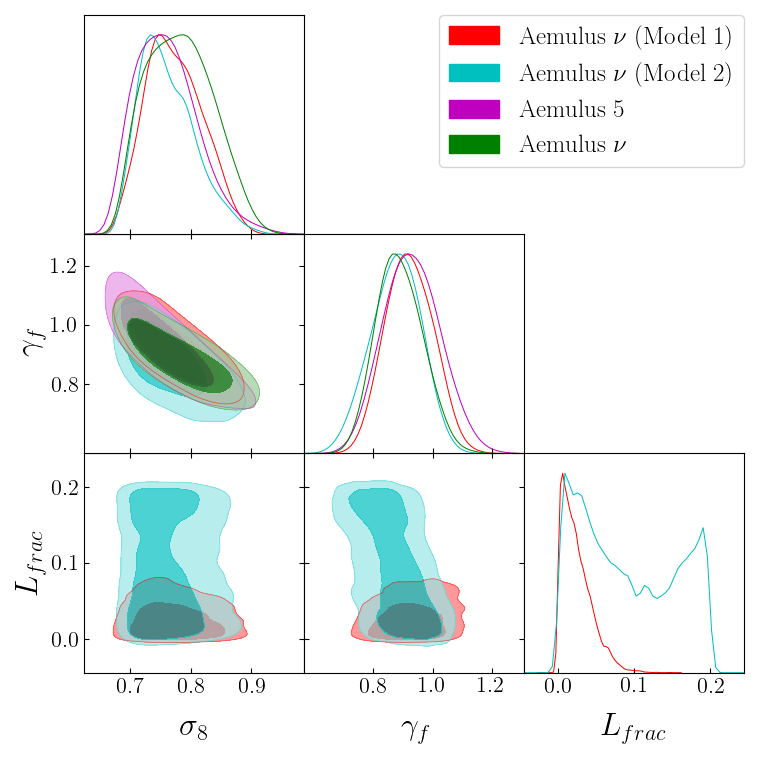}
\includegraphics[width=8cm]{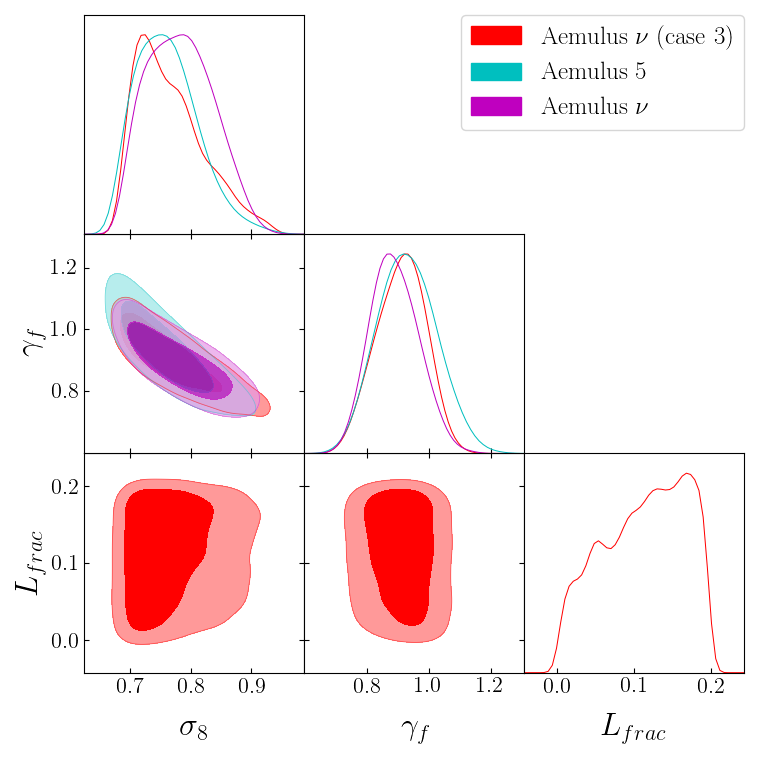}
\includegraphics[width=8cm]{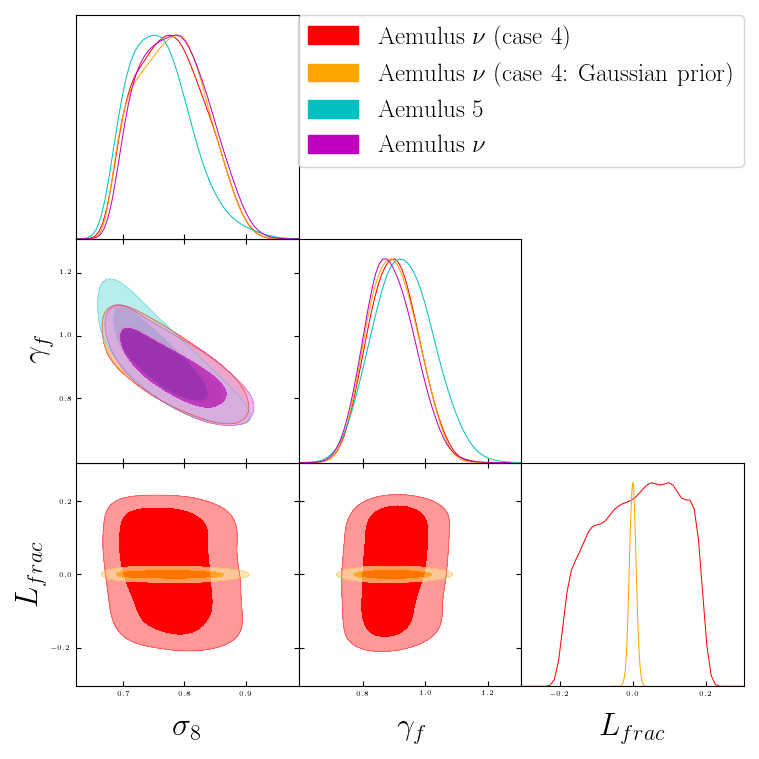}
\caption{Constraint on the growth-related parameters and the incompleteness parameter $L_{frac}$ for different models. Top left: case 1; Top right: case 2; Bottom left: case 3; Bottom right: case 4.  The result from Aemulus 5 and Aemulus $\nu$ is also shown for reference. More details are provided in the text. }
\label{fig:orphan_constraint}
\end{center}
\end{figure*}

\subsubsection{case 1}

The first case is a minimal extension of the model. Using the HOD mocks produced with the Aemulus $\nu$ suite, we randomly select a fraction $L_{frac}$ of galaxies and remove them from the catalog. To preserve the number density of the mock galaxies used to generate the HOD mocks, we artificially add $L_{frac}$ galaxies back to the catalog, but with some changes. For simplicity, we denote these galaxies as lost galaxies, noting that this terminology may be similar to "orphan" galaxies widely used in literature, but we use a different name as the situations are different. We consider two methods for the distribution of these lost galaxies in phase space. The first method assumes that these lost galaxies are randomly distributed in space and are essentially unclustered, acting as an extreme case. The second method is similar, but we use a random subsample of the dark matter particles to assign both positions and velocities to the lost galaxies.

\subsubsection{case 2}

Our HOD model assumes that all the mocks are generated with the same number density. For the analysis of Aemulus V at $z=0.55$, it is $2\times10^{-4}[h^{-1}\text{Mpc}]^{-3}$. This number density is used as input to determine the HOD parameter $M_{\text{min}}$. This means that the galaxy mocks in case 1 correspond to an HOD model with a number density $L_{frac}$ lower than the input. To correct this issue, we propose a second case where, when we generate the HOD mocks, the input number density is obtained by scaling the fiducial one by $(1-L_{frac})$. This results in an HOD mock with a number density lower than expected. Next, we apply the same two methods as in case 1 to add extra $L_{frac}$ galaxies to the mock. The resultant mock still has the number density as the fiducial one. This requirement should be fulfilled since the observed sample can have the number density well measured.

\subsubsection{case 3}

This case is simple and considers the opposite situation, where we scale the fiducial number density by a factor ($1+L_{frac}$) and re-generate the corresponding HOD mocks. Then we randomly pick $L_{frac}$ galaxies and remove them from the catalog with a remaining subsample that can still match the required number density, implying that $L_{frac}$ galaxies are simply lost from the sample. 

\subsubsection{case 4}

The above cases all deal with adding/removing a fraction of galaxies but differ in how to match the number density. In this case, the simplest model can just assume the number density as a free parameter. To avoid confusion caused by different symbols, we model this effect using the $L_{frac}$ parameter as a fractional change with respect to the required input number density.

\subsubsection{Results}

With the above models for different considerations, we re-generate the HOD mocks in the extended parameter space with $L_{frac}$ and rerun the cosmological constraint on the CMASS galaxies in Aemulus V. In Figure~\ref{fig:orphan_constraint}, we present the constraints for each model. For reference, the original constraints from Aemulus V and the re-analysis using Aemulus $\nu$ are also shown. In the first three cases, the range of $L_{frac}$ is from 0 to 0.2, while in the last case $L_{frac}$ can vary from -0.2 to 0.2.

For the growth-related parameters, we can see that the new modeling of $L_{frac}$ doesn't show a significant impact on the results, indicating that the incompleteness of the modeling does not yield a biased measurement of $f\sigma_{8}$. The first two cases show quite similar constraints on the incompleteness parameter $L_{frac}$. With the first method of adding galaxies that are not clustered, the constraint on $L_{frac}$ is up to a few percent level. On the other hand, the second method of adding galaxies by dark matter particles allows a much wider range of $L_{frac}$. This is not surprising since the first method is an extreme model to artificially exclude galaxy pairs in the correlation function, and the impact is much more radical than the second method which still uses the clustered dark matter particles.

In the opposite model of case 3, the constraint on $L_{frac}$ is weak and dominated by the prior. Similar to the previous two cases, the additional degree of freedom does not degrade the cosmological constraint either. Part of the reason is that the change of number density up to 20\% can only slightly change the halo mass of the galaxy sample given the Schechter form of the halo mass function. The impact can be much smaller than 0.1 dex on the halo mass range without a significant change in the galaxy bias.

The last case, assuming number density as a free parameter, can be an important test for the parameter degeneracy since it has been fixed in our previous works. The reason is that fixing the number density is equivalent to bringing the abundance information to anchor the average halo mass of the sample and thus the galaxy bias. This may translate to the constraint on the cosmological parameters as well. Therefore, to assess the possible impact, we construct HOD mocks and build emulators with number density as a free parameter. Note that this parameter is expressed as a fractional change with respect to the required one. We find that the cosmological constraint is quite stable, but the measurement on $L_{frac}$ itself is not tight and can be dominated by the prior. 

We also perform another test which assumes a 1\% error in the measured number density. We add this information as a Gaussian prior and rerun the test. This is similar to the test in the early BOSS LOWZ clustering analysis (\citealt{Parejko_LOWZ}) which assumes a 15\% error on the galaxy number density, but this level of uncertainty corresponds to the variation of number density as a function of redshift. Our selection of BOSS galaxies can have a much lower error. In \cite{Storey-Fisher_2022}, we use the BOSS QPM mocks and estimate that the fractional change of number density is 0.43\%. Therefore, our assumption of 1\% in the analysis is a conservative choice. For the constraint on growth-related parameters, we find that it is consistent with the fiducial one when $L_{frac}$ is free. This consistency seems to imply that most of the cosmological information is from the clustering signals rather than the abundance. This also demonstrates the robustness of our approach, and fixing the number density as an external constraint doesn't significantly degrade the cosmological measurements.

\begin{figure}
\begin{center}
\includegraphics[width=8cm]{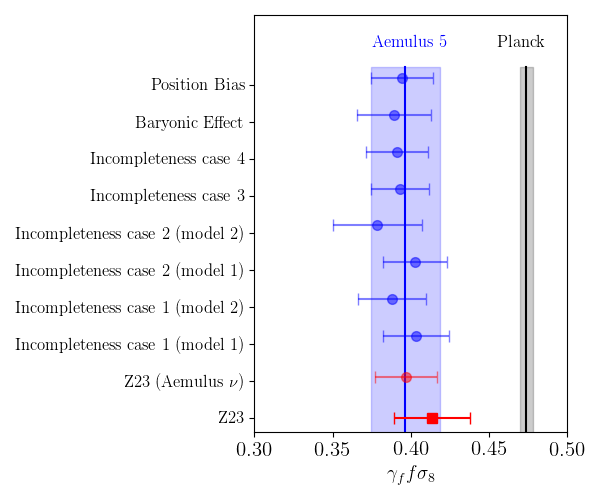}
\caption{Measurement of $f\sigma_{8}$ with different model extensions. For reference, the original result from Zhai et al. (2023a) is shown as red square, a re-analysis using Aemulus $\nu$ with assembly bias turned off is shown as red dot. The results show that model incompleteness, baryonic effect and position bias for central galaxies as model extensions of the HOD framework doesn't have significant impact on the measurement of cosmological growth. The result using galaxy samples selected by other properties is similar.}
\label{fig:orphan_constraint_growth}
\end{center}
\end{figure}

Next we apply this new model with extensions of model incompleteness to the BOSS galaxies split by different properties as in Z23. Among these multiple combinations of data and models, we do not find significant deviation in the measurement of linear growth rate. In Figure~\ref{fig:orphan_constraint_growth}, we show an example when the sample is selected by luminosity. It shows that adding the new degrees of freedom for sample incompleteness into the HOD approach doesn't strongly affect the cosmological measurements.

\subsection{Baryonic effects}\label{sec:baryon}

The clustering analysis conducted in this study relies on gravity-only simulations. While we have employed a flexible HOD model to capture the relationship between galaxies and dark matter halos, effectively accounting for the influence of baryonic physics at small scales, it is likely that this model does not fully address the uncertainties associated with galaxy formation physics. Given this limitation, various approaches have been proposed and implemented to investigate the modeling and implications within the cosmological framework. For instance the BACCO simulation employs numerical simulation and machine learning to quantify the impact of baryonic effect on the matter power spectrum (\citealt{Arico_2020}), \cite{Chaves-Montero_2023} explicitly explore the potential of resolving the $S_{8}$ tension with galaxy formation models and \cite{Kwan_2023b} employs the BAHAMA simulation to study the baryonic effect on galaxy clustering at small scales.

In this section, we take a step in a similar direction but without explicitly delving into complex baryonic models involving star formation, gas cooling, or AGN feedback. Instead, we introduce a simple single-parameter extension of the HOD model. To incorporate the effects of approximate baryonic processes on the output of gravity-only simulations, we introduce a new parameter $B_{f}$ that scales the dark matter halo mass. We then use the scaled mass to generate the HOD mocks. We artificially set the range of this new parameter to be [0.5, 1.5] and re-run the analysis, including the generation of mocks and construction of the emulator. We find that this new parameter does not significantly degrade the accuracy of the clustering statistics as we initially anticipated. Subsequently, we apply the model to the BOSS CMASS galaxies to examine the impact on cosmological measurements, similar to the approach in previous sections.

\begin{figure}
\begin{center}
\includegraphics[width=8cm]{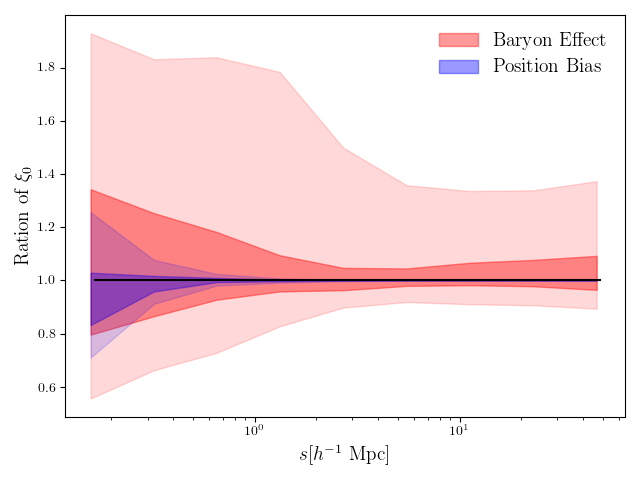}
\caption{Ratio of $\xi_{0}$ in redshift space considering the effect from baryonic effect (red) and position bias of central galaxies (blue). The results come from the Aemulus $\nu$ tier 2 simulations with hundreds of HOD models randomly distributed in the parameter space, expressed as the ratio compared with the results without corrections from baryonic effect or position bias of centrals. The shaded area denotes the inner 68\% and 95\% distribution. The behavior for $w_{p}$ and $\xi_{2}$ is similar.}
\label{fig:baryon_xi0_effect}
\end{center}
\end{figure}

We first examine the impact of this new parameter on the galaxy correlation function using the Aemulus $\nu$ tier 2 simulations. In Figure~\ref{fig:baryon_xi0_effect}, we present the result for $\xi_{0}$ in redshift space by computing the ratio with and without the $B_{f}$ parameter for hundreds of HOD models. The red shaded area shows the inner 68\% and 95\% distribution of the samples. The result shows that the change in halo mass of up to 50\% can have a significant impact on the amplitude of the correlation function, with a larger impact at small scales where the one-halo term is more dominant.

\begin{figure*}
\begin{center}
\includegraphics[width=8cm]{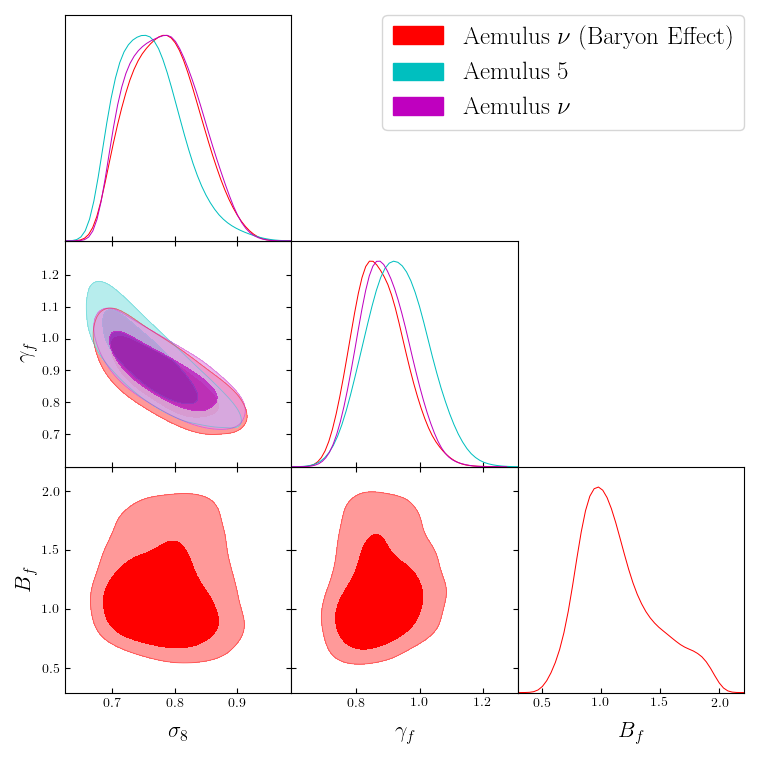}
\includegraphics[width=8cm]{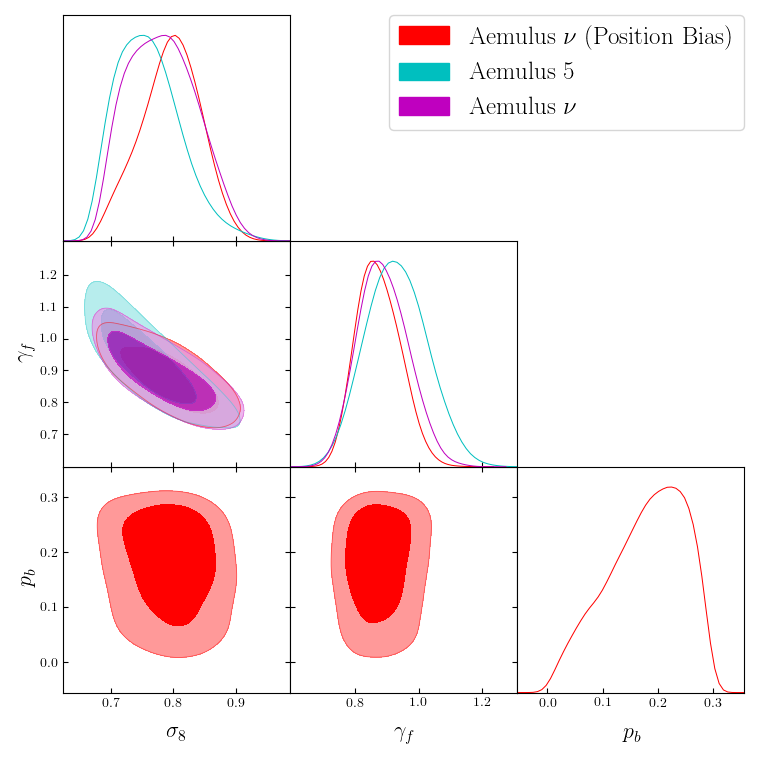}
\caption{Similar to Figure~\ref{fig:orphan_constraint} but for baryonic model (left) and position bias of central galaxies (right). The results from Aemulus 5 and Aemulus $\nu$ are also shown for reference.}
\label{fig:baryon_constraint}
\end{center}
\end{figure*}

\begin{figure}
\begin{center}
\includegraphics[width=8cm]{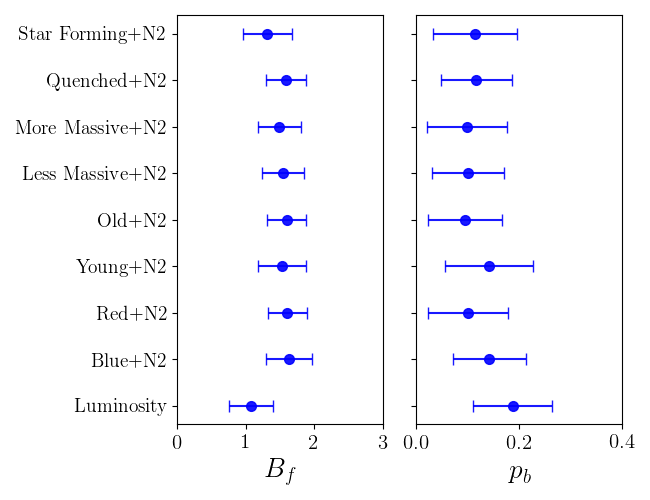}
\caption{Constraint on the extended model parameter from the BOSS galaxies split by different properties. {\it Left:} baryonic parameter $B_{f}$; {\it Right:} position bias of central galaxies.}
\label{fig:baryon_parameter_dis}
\end{center}
\end{figure}

In the left-hand panel of Figure~\ref{fig:baryon_constraint}, we show the constraint on the growth rate related parameters considering the baryonic effect and compare them with the reference results. It shows that the cosmological inference is almost unchanged, indicating that the approximate model doesn't affect the measurement of $f\sigma_{8}$. On the other hand, the constraint on the parameter $B_{f}$ is centered at unity. We do not see a noticeable baryonic effect from this clustering analysis. In order to see if this result is due to a statistical fluke or not, we also run tests with the new emulator on the CMASS galaxies split by different properties and find that although the growth rate measurement is not significantly affected (Figure~\ref{fig:orphan_constraint_growth}), the constraint on $B_{f}$ is somewhat higher than Aemulus V. We display the constraint in the left panel of Figure~\ref{fig:baryon_parameter_dis}. Because $B_{f}$ is a scaling factor for halo mass, the preference for $B_{f}>1$ indicates that the halo mass prediction from the N-body simulation is somewhat underestimated. However, the deviation from unity is less than $2\sigma$, and thus the finding is not conclusive. In addition, $B_{f}$ is correlated with other parameters, affecting both centrals and satellites. The high value of $B_{f}$ correlates with slightly lower $f_{max}$, the incompleteness parameter of central occupancy at the massive end, and lower $\eta_{vs}$, the velocity bias of satellites and slightly higher $\alpha$ for satellite occupancy. In general, a high $B_{f}$ can produce a galaxy mock with higher satellite fraction and thus lower number of centrals in order to preserve the galaxy number density. To maintain the galaxy bias anchored by a large scale measurement, the lower number of centrals has to be offset by less occupancy of the more massive halos and thus a lower value of $f_{max}$. Since the overall impact of $B_{f}$ is relatively weak at large scales where the centrals dominate (Fig.~\ref{fig:baryon_xi0_effect}), this influence is not very significant. For the same reason, a high value of $B_{f}$ can limit satellites to a narrower range of halo mass (a halo before scaling without satellites can host satellites after scaling, and there are fewer halos available at the massive end due to $f_{max}$). The model has to produce more satellite pairs with a faster occupancy (higher $\alpha$) as a function of halo mass. In order to balance the enhanced clustering due to more one-halo terms, the velocity field needs to be suppressed to fit the data.
 
This interpretation does not consider effects on the internal structures of the dark matter halos such as the radius or radial density profile. These may partially explain the constraint on $B_{f}$ since the cosmological measurement from clustering also depends on the velocity field, which is not modified in our model. Additionally, the change in dark matter halo mass behaves exactly as that due to assembly bias, but our model is more simplified and can absorb any dependency on parameters rather than halo mass. A more realistic approach could mimic models like the decorated HOD that can redistribute the halo catalog during mock construction as presented in  \citet{Hearin_2016}.

\subsection{Position bias of central galaxies}\label{sec:central}

In the HOD model, we assume that central galaxies reside at the centers of their dark matter halos. In reality, however, the spatial distribution of central galaxies within their halos can be complex, influenced by factors such as halo mergers, interactions with neighboring galaxies, and the local cosmological environment. Therefore, the assumption that central galaxies are located at the centers of their halos should be reexamined. This effect is particularly important in the study of galaxy clusters and weak lensing, as misidentifying the cluster center can significantly impact the modeling and understanding of halo dynamics (\citealt{Yang_2006, Johnston_2007}). To differentiate this effect from the mis-centering effect in galaxy clusters, we simply refer to it as "position bias" for central galaxies in our analysis.

We relax this assumption by introducing a new parameter $p_{b}$. For central galaxies in the mock catalogs, we first identify their host halos and halo radius $R_{h}$. We then multiply $R_{h}$ by $p_{b}$ and apply a Gaussian dispersion with a mean of zero and a width of $p_{b}R_{h}$. Next, we draw random values from this distribution and add them as perturbations to the positions of the central galaxies. In this way, $p_{b}$ describes the strength of the offset of central galaxies compared to the halo center, and we choose a range of [0, 0.3] for $p_{b}$ in our modeling.

Similar to the previous test on the baryonic effect, we display the impact on $\xi_{0}$ in Figure~\ref{fig:baryon_xi0_effect}. We can see that the overall impact is less significant than the baryonic effect, but this is mainly due to the range of $p_{b}$ we have considered. On the other hand, the position bias is more important at small scales and has almost no impact at large scales. This is consistent with expectations, as at large scales, the position offset of central galaxies is negligible compared to the distances between galaxies. For the mass range of interest, the average perturbation in distance is around a few tens to hundreds of kpc. However, this change is not negligible for galaxy pairs within massive halos, thus it can alter the correlation function in the one-halo regime. This is interesting because the model only applies to central galaxies but has a more significant effect at small scales.

With this extended model, we make cosmological constraints using the CMASS galaxies and present the results in the right-hand panel of Figure~\ref{fig:baryon_constraint}. Similar to previous work, we do not observe a clear impact on the measurement of the linear growth rate from this position bias. However, we find that the results favor a non-zero value of $p_{b}$ and peak around 0.2. This suggests that central galaxies have a preference for not residing in the center of their host halos. We also conduct the same tests on the BOSS galaxies split by different properties (Figure~\ref{fig:orphan_constraint_growth}). The right panel of Figure~\ref{fig:baryon_parameter_dis} shows the constraint on $p_{b}$. The results are similar to the tests on the baryonic model, as these subsamples provide slightly different constraints compared to Aemulus V. Their constraint on $p_{b}$ is closer to 0, indicating no or weak offset in the positions of central galaxies.

\section{Discussion and Conclusion}

The observational data from state-of-the-art sky surveys have enabled accurate measurements of small-scale galaxy clustering. These measurements contain a great amount of information about the underlying cosmology and the physics governing galaxy formation and evolution but, an unbiased retrieval of this information, requires a robust understanding of the correlation between galaxies and their host dark matter halos. Due to its simplicity and economy, the HOD model has been widely used for this purpose. Its construction for different types of galaxies has provided useful information about how galaxies evolve over a wide range of redshift. On the other hand, we should be concerned that its basic assumptions and prescriptions may have a direct impact on the cosmological measurements and the ignorance of any component may have a noticeable effect on the cosmological interpretation. This limitation can be somewhat examined and mitigated by introducing additional degrees of freedom. And we have witnessed this development in the past decades with the basic HOD model extended for various purposes. In this work, we continue in this direction by introducing additional model parameters and explicitly examine their impact on the cosmological measurements with the BOSS galaxy sample using both the Aemulus (\citealt{DeRose_2018}) and Aemulus $\nu$ (\citealt{DeRose_2023}) simulation suites. 

As an empirical model, the extension of the HOD model can be described using simple parameterizations motivated by physical numerical simulations or observational requirements (\citealt{Beltz-Mohrmann_2020, Beltz-Mohrmann_2023}). We consider the spatial distribution of satellite galaxies within the host halo and that, because of baryonic effects, the halo mass can be different from a dark matter only simulation. For the latter, we consider that central galaxy positions can be offset from the center of the host halos, a phenomena similar to the miscentering effect observed in the study of galaxy clusters. In addition, we consider the effect of sample incompleteness. We implement these extended HOD models using numerical simulations showing how they impact the correlation function as a function of scale in both real and redshift space. With the new model parameters, we adopt our previously developed emulator approach to retrieve cosmological constraints and isolate the influence from these new parameters. Interestingly, the results from the BOSS galaxies show that these new parameters do not affect the cosmological constraint significantly in terms of the measurement of linear growth rate $f\sigma_{8}$. 

The robustness of the cosmological constraints from information at small scales is interesting, and it is good to note that the basic HOD model is still sufficient to retrieve the key information about the underlying cosmology without considering additional complexities. In essence, the new degrees of freedom do not have a significant correlation with the underlying cosmology or RSD measurements (e.g. \citealt{Padilla_2019, Contreras_2021, Alimi_2024}). Our analysis is limited by the galaxy sample considered in terms of its sample variance and shot noise, and the summary statistics we use. More data would provide better constraints on the extensions of our model and rule out possible bias of the problem. In this case, the future work will be necessary to pin down the behavior of these new degrees of freedom.

The current BOSS galaxies cover a cosmic volume comparable to the typical simulation box of Aemulus or Aemulus $\nu$. This is sufficient for a small scale analysis like that we have undertaken. However, the number density only allows observations of bright galaxies and this can lead to significant noise at small scales. At scales below a few $h^{-1}$Mpc, it's likely that the cosmological information is contaminated by both the shot noise and complicated baryonic physics. We have seen such behavior in the scale dependent analysis (e.g. Figure 10 of Aemulus V): adding small scales into the analysis doesn't improve the cosmological constraints. Our updated analysis in this work also reveals similar findings. The shape modeling for satellites and position bias for central galaxies only affects clustering in the one-halo regime, and these new parameters are not strongly correlated with cosmological parameters. Given the uncertainties of the constraints, it could be useful to add more statistics in addition to the standard two-point correlation function (\citealt{Hahn_2020, Hahn_2024}), to break these degeneracies. But this risks introducing further complexities in the modelling, and we leave such investigation in future works.

In addition to the cosmological measurements, we also constrain the parameters quantifying the model extensions. We don't observe significant constraints on these parameters or evidence for strong deviations from the basic HOD model. However, we do see some weak to mild constraints for a subset of these models. For instance, our analysis for the baryonic effect shows that the halo mass from the dark matter only simulation is somewhat underestimated, and the central galaxies have a preference of not residing in the center of the host halo. Although the deviations from the standard HOD model are not strong, it is worthwhile to revisit and exploit with future and better data. From this point of view, the model and methodology developed in this work can be considered a pilot study for ongoing and future sky surveys such as DESI. The physics of the tracers observed in the future will be very different and the resultant empirical modeling will have to vary significantly. Therefore our extended model can provide the additional flexibility to describe the clustering properties of these targets and build an unbiased connection with dark matter halos. 

Since the extended parameters can change the clustering properties of galaxies, we can understand these models within the framework of assembly bias or secondary bias. However, the approach we adopt is quite different from the literature, e.g. \cite{Xu_2020}. We could do a similar analysis by looking at samples split by different properties. Then the comparison of the clustering measurement would provide an empirical prescription of the assembly bias parameters to augment the basic HOD model (e.g. \citealt{Hadzhiyska_2023}). The models we have considered in this work are only a limited number of trials to improve the modeling and characterization of galaxy halo connection on small scales. There are plenty of possibilities that can be incorporated in the model, such as influence from halo spin, age, galactic conformity, galaxy and halo orientation, and many others. Our work only serves as a first step along this direction. What is clear is that the combination with the emulator approach is able to marginalize over the cosmological dependency and improve the robustness of the results.

% End of mnras_template.tex

\section*{Data Availability}

The galaxy mocks and data are available upon reasonable request to the author.

\section*{Acknowledgements}

We thank the anonymous reviewers for the helpful and insightful comments and suggestions that have significantly improved this paper. ZZ appreciates helpful comments and discussions with Renyue Cen, Xiaodong Li, Yi Zheng, Jeremy Tinker and the Aemulus team. ZZ is supported by NSFC (12373003), the National Key\&D Program of China (2023YFA1605600), and acknowledges the generous sponsorship from Yangyang Development Fund.

WP acknowledges the support of the Canadian Space Agency and the Natural Sciences and Engineering Research Council of Canada (NSERC), [funding reference number RGPIN-2019-03908].

Research at Perimeter Institute is supported in part by the Government of Canada through the Department of Innovation, Science and Economic Development Canada and by the Province of Ontario through the Ministry of Colleges and Universities. 

This research was enabled in part by support provided by Compute Ontario (computeontario.ca) and the Digital Research Alliance of Canada (alliancecan.ca).

\rm{Software:} Python,
Matplotlib \citep{matplotlib},
NumPy \citep{numpy},
SciPy \citep{scipy}.

\appendix

\bibliographystyle{mnras}
\bibliography{emu_gc_bib,software}

\begin{thebibliography}{}
\makeatletter
\relax
\def\mn@urlcharsother{\let\do\@makeother \do\$\do\&\do\#\do\^\do\_\do\%\do\~}
\def\mn@doi{\begingroup\mn@urlcharsother \@ifnextchar [ {\mn@doi@}
  {\mn@doi@[]}}
\def\mn@doi@[#1]#2{\def\@tempa{#1}\ifx\@tempa\@empty \href
  {http://dx.doi.org/#2} {doi:#2}\else \href {http://dx.doi.org/#2} {#1}\fi
  \endgroup}
\def\mn@eprint#1#2{\mn@eprint@#1:#2::\@nil}
\def\mn@eprint@arXiv#1{\href {http://arxiv.org/abs/#1} {{\tt arXiv:#1}}}
\def\mn@eprint@dblp#1{\href {http://dblp.uni-trier.de/rec/bibtex/#1.xml}
  {dblp:#1}}
\def\mn@eprint@#1:#2:#3:#4\@nil{\def\@tempa {#1}\def\@tempb {#2}\def\@tempc
  {#3}\ifx \@tempc \@empty \let \@tempc \@tempb \let \@tempb \@tempa \fi \ifx
  \@tempb \@empty \def\@tempb {arXiv}\fi \@ifundefined
  {mn@eprint@\@tempb}{\@tempb:\@tempc}{\expandafter \expandafter \csname
  mn@eprint@\@tempb\endcsname \expandafter{\@tempc}}}

\bibitem[\protect\citeauthoryear{{Abazajian} et~al.,}{{Abazajian}
  et~al.}{2009}]{Abazajian_2009}
{Abazajian} K.~N.,  et~al., 2009, \mn@doi [\apjs]
  {10.1088/0067-0049/182/2/543}, \href
  {http://adsabs.harvard.edu/abs/2009ApJS..182..543A} {182, 543}

\bibitem[\protect\citeauthoryear{{Abdalla} et~al.,}{{Abdalla}
  et~al.}{2022}]{Abdalla_2022}
{Abdalla} E.,  et~al., 2022, \mn@doi [Journal of High Energy Astrophysics]
  {10.1016/j.jheap.2022.04.002}, \href
  {https://ui.adsabs.harvard.edu/abs/2022JHEAp..34...49A} {34, 49}

\bibitem[\protect\citeauthoryear{{Alcock} \& {Paczynski}}{{Alcock} \&
  {Paczynski}}{1979}]{Alcock_1979}
{Alcock} C.,  {Paczynski} B.,  1979, \mn@doi [\nat] {10.1038/281358a0}, \href
  {https://ui.adsabs.harvard.edu/abs/1979Natur.281..358A} {281, 358}

\bibitem[\protect\citeauthoryear{{Ali-Ha{\"\i}moud} \&
  {Bird}}{{Ali-Ha{\"\i}moud} \& {Bird}}{2013}]{Alihamoud_2013}
{Ali-Ha{\"\i}moud} Y.,  {Bird} S.,  2013, \mn@doi [\mnras]
  {10.1093/mnras/sts286}, \href
  {https://ui.adsabs.harvard.edu/abs/2013MNRAS.428.3375A} {428, 3375}

\bibitem[\protect\citeauthoryear{{Alimi} \& {Koskas}}{{Alimi} \&
  {Koskas}}{2024}]{Alimi_2024}
{Alimi} J.-M.,  {Koskas} R.,  2024, \mn@doi [arXiv e-prints]
  {10.48550/arXiv.2406.15947}, \href
  {https://ui.adsabs.harvard.edu/abs/2024arXiv240615947A} {p. arXiv:2406.15947}

\bibitem[\protect\citeauthoryear{{Allgood}, {Flores}, {Primack}, {Kravtsov},
  {Wechsler}, {Faltenbacher}  \& {Bullock}}{{Allgood}
  et~al.}{2006}]{Allgood_2006}
{Allgood} B.,  {Flores} R.~A.,  {Primack} J.~R.,  {Kravtsov} A.~V.,  {Wechsler}
  R.~H.,  {Faltenbacher} A.,   {Bullock} J.~S.,  2006, \mn@doi [\mnras]
  {10.1111/j.1365-2966.2006.10094.x}, \href
  {https://ui.adsabs.harvard.edu/abs/2006MNRAS.367.1781A} {367, 1781}

\bibitem[\protect\citeauthoryear{{Aric{\`o}}, {Angulo}, {Contreras},
  {Ondaro-Mallea}, {Pellejero-Iba{\~n}ez}  \& {Zennaro}}{{Aric{\`o}}
  et~al.}{2020}]{Arico_2020}
{Aric{\`o}} G.,  {Angulo} R.~E.,  {Contreras} S.,  {Ondaro-Mallea} L.,
  {Pellejero-Iba{\~n}ez} M.,   {Zennaro} M.,  2020, arXiv e-prints, \href
  {https://ui.adsabs.harvard.edu/abs/2020arXiv201115018A} {p. arXiv:2011.15018}

\bibitem[\protect\citeauthoryear{{Banerjee} \& {Dalal}}{{Banerjee} \&
  {Dalal}}{2016}]{Banerjee_2016}
{Banerjee} A.,  {Dalal} N.,  2016, \mn@doi [\jcap]
  {10.1088/1475-7516/2016/11/015}, \href
  {https://ui.adsabs.harvard.edu/abs/2016JCAP...11..015B} {2016, 015}

\bibitem[\protect\citeauthoryear{{Banerjee}, {Powell}, {Abel}  \&
  {Villaescusa-Navarro}}{{Banerjee} et~al.}{2018}]{Banerjee_2018}
{Banerjee} A.,  {Powell} D.,  {Abel} T.,   {Villaescusa-Navarro} F.,  2018,
  \mn@doi [\jcap] {10.1088/1475-7516/2018/09/028}, \href
  {https://ui.adsabs.harvard.edu/abs/2018JCAP...09..028B} {2018, 028}

\bibitem[\protect\citeauthoryear{{Bautista} et~al.,}{{Bautista}
  et~al.}{2021}]{Bautista_2021}
{Bautista} J.~E.,  et~al., 2021, \mn@doi [\mnras] {10.1093/mnras/staa2800},
  \href {https://ui.adsabs.harvard.edu/abs/2021MNRAS.500..736B} {500, 736}

\bibitem[\protect\citeauthoryear{{Behroozi}, {Wechsler}  \& {Wu}}{{Behroozi}
  et~al.}{2013}]{Behroozi_2013b}
{Behroozi} P.~S.,  {Wechsler} R.~H.,   {Wu} H.-Y.,  2013, \mn@doi [\apj]
  {10.1088/0004-637X/762/2/109}, \href
  {http://adsabs.harvard.edu/abs/2013ApJ...762..109B} {762, 109}

\bibitem[\protect\citeauthoryear{{Beltz-Mohrmann}, {Berlind}  \&
  {Szewciw}}{{Beltz-Mohrmann} et~al.}{2020}]{Beltz-Mohrmann_2020}
{Beltz-Mohrmann} G.~D.,  {Berlind} A.~A.,   {Szewciw} A.~O.,  2020, \mn@doi
  [\mnras] {10.1093/mnras/stz3442}, \href
  {https://ui.adsabs.harvard.edu/abs/2020MNRAS.491.5771B} {491, 5771}

\bibitem[\protect\citeauthoryear{{Beltz-Mohrmann}, {Szewciw}, {Berlind}  \&
  {Sinha}}{{Beltz-Mohrmann} et~al.}{2023}]{Beltz-Mohrmann_2023}
{Beltz-Mohrmann} G.~D.,  {Szewciw} A.~O.,  {Berlind} A.~A.,   {Sinha} M.,
  2023, \mn@doi [\apj] {10.3847/1538-4357/acc576}, \href
  {https://ui.adsabs.harvard.edu/abs/2023ApJ...948..100B} {948, 100}

\bibitem[\protect\citeauthoryear{{Bett}, {Eke}, {Frenk}, {Jenkins}, {Helly}  \&
  {Navarro}}{{Bett} et~al.}{2007}]{Bett_2007}
{Bett} P.,  {Eke} V.,  {Frenk} C.~S.,  {Jenkins} A.,  {Helly} J.,   {Navarro}
  J.,  2007, \mn@doi [\mnras] {10.1111/j.1365-2966.2007.11432.x}, \href
  {https://ui.adsabs.harvard.edu/abs/2007MNRAS.376..215B} {376, 215}

\bibitem[\protect\citeauthoryear{{Bonamigo}, {Despali}, {Limousin}, {Angulo},
  {Giocoli}  \& {Soucail}}{{Bonamigo} et~al.}{2015}]{Bonamigo_2015}
{Bonamigo} M.,  {Despali} G.,  {Limousin} M.,  {Angulo} R.,  {Giocoli} C.,
  {Soucail} G.,  2015, \mn@doi [\mnras] {10.1093/mnras/stv417}, \href
  {https://ui.adsabs.harvard.edu/abs/2015MNRAS.449.3171B} {449, 3171}

\bibitem[\protect\citeauthoryear{{Brandbyge} \& {Hannestad}}{{Brandbyge} \&
  {Hannestad}}{2009}]{Brandbyge_2009}
{Brandbyge} J.,  {Hannestad} S.,  2009, \mn@doi [\jcap]
  {10.1088/1475-7516/2009/05/002}, \href
  {https://ui.adsabs.harvard.edu/abs/2009JCAP...05..002B} {2009, 002}

\bibitem[\protect\citeauthoryear{{Brown} et~al.,}{{Brown}
  et~al.}{2008}]{Brown_2008}
{Brown} M. J.~I.,  et~al., 2008, \mn@doi [\apj] {10.1086/589538}, \href
  {https://ui.adsabs.harvard.edu/abs/2008ApJ...682..937B} {682, 937}

\bibitem[\protect\citeauthoryear{{Bryan}, {Kay}, {Duffy}, {Schaye}, {Dalla
  Vecchia}  \& {Booth}}{{Bryan} et~al.}{2013}]{Bryan_2013}
{Bryan} S.~E.,  {Kay} S.~T.,  {Duffy} A.~R.,  {Schaye} J.,  {Dalla Vecchia} C.,
    {Booth} C.~M.,  2013, \mn@doi [\mnras] {10.1093/mnras/sts587}, \href
  {https://ui.adsabs.harvard.edu/abs/2013MNRAS.429.3316B} {429, 3316}

\bibitem[\protect\citeauthoryear{{Buchner} et~al.,}{{Buchner}
  et~al.}{2014}]{Buchner_2014}
{Buchner} J.,  et~al., 2014, \mn@doi [\aap] {10.1051/0004-6361/201322971},
  \href {https://ui.adsabs.harvard.edu/abs/2014A&A...564A.125B} {564, A125}

\bibitem[\protect\citeauthoryear{{Cataldi} et~al.,}{{Cataldi}
  et~al.}{2023}]{Cataldi_2023}
{Cataldi} P.,  et~al., 2023, \mn@doi [\mnras] {10.1093/mnras/stad1601}, \href
  {https://ui.adsabs.harvard.edu/abs/2023MNRAS.523.1919C} {523, 1919}

\bibitem[\protect\citeauthoryear{{Chapman} et~al.,}{{Chapman}
  et~al.}{2021}]{Chapman_2021}
{Chapman} M.~J.,  et~al., 2021, arXiv e-prints, \href
  {https://ui.adsabs.harvard.edu/abs/2021arXiv210614961C} {p. arXiv:2106.14961}

\bibitem[\protect\citeauthoryear{{Chapman}, {Zhai}  \& {Percival}}{{Chapman}
  et~al.}{2023}]{Chapman_2023}
{Chapman} M.~J.,  {Zhai} Z.,   {Percival} W.~J.,  2023, \mn@doi [\mnras]
  {10.1093/mnras/stad2351}, \href
  {https://ui.adsabs.harvard.edu/abs/2023MNRAS.tmp.2289C} {}

\bibitem[\protect\citeauthoryear{{Chaves-Montero}, {Angulo}  \&
  {Contreras}}{{Chaves-Montero} et~al.}{2023}]{Chaves-Montero_2023}
{Chaves-Montero} J.,  {Angulo} R.~E.,   {Contreras} S.,  2023, \mn@doi [\mnras]
  {10.1093/mnras/stad243}, \href
  {https://ui.adsabs.harvard.edu/abs/2023MNRAS.521..937C} {521, 937}

\bibitem[\protect\citeauthoryear{{Chen}, {Mo}, {Li}, {Wang}, {Yang}, {Zhang}
  \& {Wang}}{{Chen} et~al.}{2020}]{Chen_2020}
{Chen} Y.,  {Mo} H.~J.,  {Li} C.,  {Wang} H.,  {Yang} X.,  {Zhang} Y.,   {Wang}
  K.,  2020, \mn@doi [\apj] {10.3847/1538-4357/aba597}, \href
  {https://ui.adsabs.harvard.edu/abs/2020ApJ...899...81C} {899, 81}

\bibitem[\protect\citeauthoryear{{Chua}, {Pillepich}, {Vogelsberger}  \&
  {Hernquist}}{{Chua} et~al.}{2019}]{Chua_2019}
{Chua} K. T.~E.,  {Pillepich} A.,  {Vogelsberger} M.,   {Hernquist} L.,  2019,
  \mn@doi [\mnras] {10.1093/mnras/sty3531}, \href
  {https://ui.adsabs.harvard.edu/abs/2019MNRAS.484..476C} {484, 476}

\bibitem[\protect\citeauthoryear{{Clifton}, {Ferreira}, {Padilla}  \&
  {Skordis}}{{Clifton} et~al.}{2012}]{Clifton_2012}
{Clifton} T.,  {Ferreira} P.~G.,  {Padilla} A.,   {Skordis} C.,  2012, \mn@doi
  [\physrep] {10.1016/j.physrep.2012.01.001}, \href
  {https://ui.adsabs.harvard.edu/abs/2012PhR...513....1C} {513, 1}

\bibitem[\protect\citeauthoryear{{Cole} et~al.,}{{Cole}
  et~al.}{2005}]{Cole_2005}
{Cole} S.,  et~al., 2005, \mn@doi [\mnras] {10.1111/j.1365-2966.2005.09318.x},
  \href {http://adsabs.harvard.edu/abs/2005MNRAS.362..505C} {362, 505}

\bibitem[\protect\citeauthoryear{{Colless} et~al.,}{{Colless}
  et~al.}{2001}]{Colless_2001}
{Colless} M.,  et~al., 2001, \mn@doi [\mnras]
  {10.1046/j.1365-8711.2001.04902.x}, \href
  {http://adsabs.harvard.edu/abs/2001MNRAS.328.1039C} {328, 1039}

\bibitem[\protect\citeauthoryear{{Contreras}, {Chaves-Montero}, {Zennaro}  \&
  {Angulo}}{{Contreras} et~al.}{2021}]{Contreras_2021}
{Contreras} S.,  {Chaves-Montero} J.,  {Zennaro} M.,   {Angulo} R.~E.,  2021,
  \mn@doi [\mnras] {10.1093/mnras/stab2367}, \href
  {https://ui.adsabs.harvard.edu/abs/2021MNRAS.507.3412C} {507, 3412}

\bibitem[\protect\citeauthoryear{{DESI Collaboration} et~al.,}{{DESI
  Collaboration} et~al.}{2016}]{DESI_2016}
{DESI Collaboration} et~al., 2016, preprint, \href
  {http://adsabs.harvard.edu/abs/2016arXiv161100036D} {} (\mn@eprint {arXiv}
  {1611.00036})

\bibitem[\protect\citeauthoryear{{DESI Collaboration} et~al.,}{{DESI
  Collaboration} et~al.}{2024}]{DESI_2024_BAO}
{DESI Collaboration} et~al., 2024, \mn@doi [arXiv e-prints]
  {10.48550/arXiv.2404.03002}, \href
  {https://ui.adsabs.harvard.edu/abs/2024arXiv240403002D} {p. arXiv:2404.03002}

\bibitem[\protect\citeauthoryear{{Davis} \& {Peebles}}{{Davis} \&
  {Peebles}}{1983}]{Davis_1983}
{Davis} M.,  {Peebles} P.~J.~E.,  1983, \mn@doi [\apj] {10.1086/160884}, \href
  {http://adsabs.harvard.edu/abs/1983ApJ...267..465D} {267, 465}

\bibitem[\protect\citeauthoryear{{Dawson} et~al.,}{{Dawson}
  et~al.}{2013}]{Dawson_BOSS}
{Dawson} K.~S.,  et~al., 2013, \mn@doi [\aj] {10.1088/0004-6256/145/1/10},
  \href {http://adsabs.harvard.edu/abs/2013AJ....145...10D} {145, 10}

\bibitem[\protect\citeauthoryear{{Dawson} et~al.,}{{Dawson}
  et~al.}{2016}]{eBOSS_Dawson}
{Dawson} K.~S.,  et~al., 2016, \mn@doi [\aj] {10.3847/0004-6256/151/2/44},
  \href {http://adsabs.harvard.edu/abs/2016AJ....151...44D} {151, 44}

\bibitem[\protect\citeauthoryear{{DeRose} et~al.,}{{DeRose}
  et~al.}{2019}]{DeRose_2018}
{DeRose} J.,  et~al., 2019, \mn@doi [\apj] {10.3847/1538-4357/ab1085}, \href
  {https://ui.adsabs.harvard.edu/abs/2019ApJ...875...69D} {875, 69}

\bibitem[\protect\citeauthoryear{{DeRose} et~al.,}{{DeRose}
  et~al.}{2023}]{DeRose_2023}
{DeRose} J.,  et~al., 2023, \mn@doi [\jcap] {10.1088/1475-7516/2023/07/054},
  \href {https://ui.adsabs.harvard.edu/abs/2023JCAP...07..054D} {2023, 054}

\bibitem[\protect\citeauthoryear{{Doroshkevich}}{{Doroshkevich}}{1970}]{Doroshkevich_1970}
{Doroshkevich} A.~G.,  1970, \mn@doi [Astrophysics] {10.1007/BF01001625}, \href
  {https://ui.adsabs.harvard.edu/abs/1970Ap......6..320D} {6, 320}

\bibitem[\protect\citeauthoryear{{Drinkwater} et~al.,}{{Drinkwater}
  et~al.}{2010}]{Drinkwater_2010}
{Drinkwater} M.~J.,  et~al., 2010, \mn@doi [\mnras]
  {10.1111/j.1365-2966.2009.15754.x}, \href
  {http://adsabs.harvard.edu/abs/2010MNRAS.401.1429D} {401, 1429}

\bibitem[\protect\citeauthoryear{{Durkalec}, {Pollo}  \& {Abbas}}{{Durkalec}
  et~al.}{2024}]{Durkalec_2024}
{Durkalec} A.,  {Pollo} A.,   {Abbas} U.,  2024, \mn@doi [\apj]
  {10.3847/1538-4357/ad36c6}, \href
  {https://ui.adsabs.harvard.edu/abs/2024ApJ...966...73D} {966, 73}

\bibitem[\protect\citeauthoryear{{Euclid Collaboration} et~al.,}{{Euclid
  Collaboration} et~al.}{2019}]{Knabenhans_2019}
{Euclid Collaboration} et~al., 2019, \mn@doi [\mnras] {10.1093/mnras/stz197},
  \href {https://ui.adsabs.harvard.edu/abs/2019MNRAS.484.5509E} {484, 5509}

\bibitem[\protect\citeauthoryear{{Faltenbacher} \& {White}}{{Faltenbacher} \&
  {White}}{2010}]{Faltenbacher_2010}
{Faltenbacher} A.,  {White} S. D.~M.,  2010, \mn@doi [\apj]
  {10.1088/0004-637X/708/1/469}, \href
  {https://ui.adsabs.harvard.edu/abs/2010ApJ...708..469F} {708, 469}

\bibitem[\protect\citeauthoryear{{Feroz}, {Hobson}  \& {Bridges}}{{Feroz}
  et~al.}{2009}]{Feroz_2009}
{Feroz} F.,  {Hobson} M.~P.,   {Bridges} M.,  2009, \mn@doi [\mnras]
  {10.1111/j.1365-2966.2009.14548.x}, \href
  {https://ui.adsabs.harvard.edu/abs/2009MNRAS.398.1601F} {398, 1601}

\bibitem[\protect\citeauthoryear{{Ferraro}, {Sailer}, {Slosar}  \&
  {White}}{{Ferraro} et~al.}{2022}]{Ferraro_2022}
{Ferraro} S.,  {Sailer} N.,  {Slosar} A.,   {White} M.,  2022, \mn@doi [arXiv
  e-prints] {10.48550/arXiv.2203.07506}, \href
  {https://ui.adsabs.harvard.edu/abs/2022arXiv220307506F} {p. arXiv:2203.07506}

\bibitem[\protect\citeauthoryear{{Franx}, {Illingworth}  \& {de Zeeuw}}{{Franx}
  et~al.}{1991}]{Franx_1991}
{Franx} M.,  {Illingworth} G.,   {de Zeeuw} T.,  1991, \mn@doi [\apj]
  {10.1086/170769}, \href
  {https://ui.adsabs.harvard.edu/abs/1991ApJ...383..112F} {383, 112}

\bibitem[\protect\citeauthoryear{{Fraser}, {Paillas}, {Percival}, {Nadathur},
  {Radinovi{\'c}}  \& {Winther}}{{Fraser} et~al.}{2024}]{Fraser2024}
{Fraser} T.~S.,  {Paillas} E.,  {Percival} W.~J.,  {Nadathur} S.,
  {Radinovi{\'c}} S.,   {Winther} H.~A.,  2024, \mn@doi [arXiv e-prints]
  {10.48550/arXiv.2407.03221}, \href
  {https://ui.adsabs.harvard.edu/abs/2024arXiv240703221F} {p. arXiv:2407.03221}

\bibitem[\protect\citeauthoryear{{Gao} et~al.}{{Gao} et~al.}{2024}]{Gao_2024}
{Gao} W.,  et~al., 2024, in preparation

\bibitem[\protect\citeauthoryear{{Guo} et~al.,}{{Guo} et~al.}{2015}]{Guo_2015}
{Guo} H.,  et~al., 2015, \mn@doi [\mnras] {10.1093/mnras/stu2120}, \href
  {https://ui.adsabs.harvard.edu/abs/2015MNRAS.446..578G} {446, 578}

\bibitem[\protect\citeauthoryear{{Hadzhiyska} et~al.,}{{Hadzhiyska}
  et~al.}{2023}]{Hadzhiyska_2023}
{Hadzhiyska} B.,  et~al., 2023, \mn@doi [\mnras] {10.1093/mnras/stad279}, \href
  {https://ui.adsabs.harvard.edu/abs/2023MNRAS.524.2524H} {524, 2524}

\bibitem[\protect\citeauthoryear{{Hahn}, {Villaescusa-Navarro}, {Castorina}  \&
  {Scoccimarro}}{{Hahn} et~al.}{2020}]{Hahn_2020}
{Hahn} C.,  {Villaescusa-Navarro} F.,  {Castorina} E.,   {Scoccimarro} R.,
  2020, \mn@doi [\jcap] {10.1088/1475-7516/2020/03/040}, \href
  {https://ui.adsabs.harvard.edu/abs/2020JCAP...03..040H} {2020, 040}

\bibitem[\protect\citeauthoryear{{Hahn} et~al.,}{{Hahn}
  et~al.}{2024}]{Hahn_2024}
{Hahn} C.,  et~al., 2024, \mn@doi [\prd] {10.1103/PhysRevD.109.083534}, \href
  {https://ui.adsabs.harvard.edu/abs/2024PhRvD.109h3534H} {109, 083534}

\bibitem[\protect\citeauthoryear{{Han}, {Li}, {Jing}, {Nishimichi}, {Wang}  \&
  {Jiang}}{{Han} et~al.}{2019}]{Han_2019}
{Han} J.,  {Li} Y.,  {Jing} Y.,  {Nishimichi} T.,  {Wang} W.,   {Jiang} C.,
  2019, \mn@doi [\mnras] {10.1093/mnras/sty2822}, \href
  {https://ui.adsabs.harvard.edu/abs/2019MNRAS.482.1900H} {482, 1900}

\bibitem[\protect\citeauthoryear{{Hearin}, {Zentner}, {van den Bosch},
  {Campbell}  \& {Tollerud}}{{Hearin} et~al.}{2016}]{Hearin_2016}
{Hearin} A.~P.,  {Zentner} A.~R.,  {van den Bosch} F.~C.,  {Campbell} D.,
  {Tollerud} E.,  2016, \mn@doi [\mnras] {10.1093/mnras/stw840}, \href
  {https://ui.adsabs.harvard.edu/abs/2016MNRAS.460.2552H} {460, 2552}

\bibitem[\protect\citeauthoryear{{Heitmann}, {Higdon}, {White}, {Habib},
  {Williams}, {Lawrence}  \& {Wagner}}{{Heitmann} et~al.}{2009}]{Heitmann_2009}
{Heitmann} K.,  {Higdon} D.,  {White} M.,  {Habib} S.,  {Williams} B.~J.,
  {Lawrence} E.,   {Wagner} C.,  2009, \mn@doi [\apj]
  {10.1088/0004-637X/705/1/156}, \href
  {http://adsabs.harvard.edu/abs/2009ApJ...705..156H} {705, 156}

\bibitem[\protect\citeauthoryear{{Heitmann}, {White}, {Wagner}, {Habib}  \&
  {Higdon}}{{Heitmann} et~al.}{2010}]{Heitmann_2010}
{Heitmann} K.,  {White} M.,  {Wagner} C.,  {Habib} S.,   {Higdon} D.,  2010,
  \mn@doi [\apj] {10.1088/0004-637X/715/1/104}, \href
  {http://adsabs.harvard.edu/abs/2010ApJ...715..104H} {715, 104}

\bibitem[\protect\citeauthoryear{Hunter}{Hunter}{2007}]{matplotlib}
Hunter J.~D.,  2007, \mn@doi [Computing in Science Engineering]
  {10.1109/MCSE.2007.55}, 9, 90

\bibitem[\protect\citeauthoryear{{Jeeson-Daniel}, {Dalla Vecchia}, {Haas}  \&
  {Schaye}}{{Jeeson-Daniel} et~al.}{2011}]{JeesonDaniel_2011}
{Jeeson-Daniel} A.,  {Dalla Vecchia} C.,  {Haas} M.~R.,   {Schaye} J.,  2011,
  \mn@doi [\mnras] {10.1111/j.1745-3933.2011.01081.x}, \href
  {https://ui.adsabs.harvard.edu/abs/2011MNRAS.415L..69J} {415, L69}

\bibitem[\protect\citeauthoryear{{Jing} \& {Suto}}{{Jing} \&
  {Suto}}{2002}]{Jing_2002}
{Jing} Y.~P.,  {Suto} Y.,  2002, \mn@doi [\apj] {10.1086/341065}, \href
  {https://ui.adsabs.harvard.edu/abs/2002ApJ...574..538J} {574, 538}

\bibitem[\protect\citeauthoryear{{Johnston}, {Sheldon}, {Tasitsiomi},
  {Frieman}, {Wechsler}  \& {McKay}}{{Johnston} et~al.}{2007}]{Johnston_2007}
{Johnston} D.~E.,  {Sheldon} E.~S.,  {Tasitsiomi} A.,  {Frieman} J.~A.,
  {Wechsler} R.~H.,   {McKay} T.~A.,  2007, \mn@doi [\apj] {10.1086/510060},
  \href {https://ui.adsabs.harvard.edu/abs/2007ApJ...656...27J} {656, 27}

\bibitem[\protect\citeauthoryear{Jones, Oliphant, Peterson  et~al.}{Jones
  et~al.}{01  }]{scipy}
Jones E.,  Oliphant T.,  Peterson P.,   et~al., 2001--, {SciPy}: Open source
  scientific tools for {Python}, \url {http://www.scipy.org/}

\bibitem[\protect\citeauthoryear{{Kasun} \& {Evrard}}{{Kasun} \&
  {Evrard}}{2005}]{Kasun_2005}
{Kasun} S.~F.,  {Evrard} A.~E.,  2005, \mn@doi [\apj] {10.1086/430811}, \href
  {https://ui.adsabs.harvard.edu/abs/2005ApJ...629..781K} {629, 781}

\bibitem[\protect\citeauthoryear{{Klypin}, {Yepes}, {Gottl{\"o}ber}, {Prada}
  \& {He{\ss}}}{{Klypin} et~al.}{2016}]{Klypin_2016}
{Klypin} A.,  {Yepes} G.,  {Gottl{\"o}ber} S.,  {Prada} F.,   {He{\ss}} S.,
  2016, \mn@doi [\mnras] {10.1093/mnras/stw248}, \href
  {http://adsabs.harvard.edu/abs/2016MNRAS.457.4340K} {457, 4340}

\bibitem[\protect\citeauthoryear{{Kwan}, {Heitmann}, {Habib}, {Padmanabhan},
  {Lawrence}, {Finkel}, {Frontiere}  \& {Pope}}{{Kwan}
  et~al.}{2015}]{Kwan_2015}
{Kwan} J.,  {Heitmann} K.,  {Habib} S.,  {Padmanabhan} N.,  {Lawrence} E.,
  {Finkel} H.,  {Frontiere} N.,   {Pope} A.,  2015, \mn@doi [\apj]
  {10.1088/0004-637X/810/1/35}, \href
  {https://ui.adsabs.harvard.edu/\#abs/2015ApJ...810...35K} {810, 35}

\bibitem[\protect\citeauthoryear{{Kwan}, {McCarthy}  \& {Salcido}}{{Kwan}
  et~al.}{2023a}]{Kwan_2023b}
{Kwan} J.,  {McCarthy} I.~G.,   {Salcido} J.,  2023a, \mn@doi [arXiv e-prints]
  {10.48550/arXiv.2312.05009}, \href
  {https://ui.adsabs.harvard.edu/abs/2023arXiv231205009K} {p. arXiv:2312.05009}

\bibitem[\protect\citeauthoryear{{Kwan} et~al.,}{{Kwan}
  et~al.}{2023b}]{Kwan_2023}
{Kwan} J.,  et~al., 2023b, \mn@doi [\apj] {10.3847/1538-4357/acd92f}, \href
  {https://ui.adsabs.harvard.edu/abs/2023ApJ...952...80K} {952, 80}

\bibitem[\protect\citeauthoryear{{Lange}, {Hearin}, {Leauthaud}, {van den
  Bosch}, {Guo}  \& {DeRose}}{{Lange} et~al.}{2021}]{Lange_2021}
{Lange} J.~U.,  {Hearin} A.~P.,  {Leauthaud} A.,  {van den Bosch} F.~C.,  {Guo}
  H.,   {DeRose} J.,  2021, arXiv e-prints, \href
  {https://ui.adsabs.harvard.edu/abs/2021arXiv210112261L} {p. arXiv:2101.12261}

\bibitem[\protect\citeauthoryear{{Lange}, {Hearin}, {Leauthaud}, {van den
  Bosch}, {Xhakaj}, {Guo}, {Wechsler}  \& {DeRose}}{{Lange}
  et~al.}{2023}]{Lange_2023}
{Lange} J.~U.,  {Hearin} A.~P.,  {Leauthaud} A.,  {van den Bosch} F.~C.,
  {Xhakaj} E.,  {Guo} H.,  {Wechsler} R.~H.,   {DeRose} J.,  2023, \mn@doi
  [\mnras] {10.1093/mnras/stad473}, \href
  {https://ui.adsabs.harvard.edu/abs/2023MNRAS.520.5373L} {520, 5373}

\bibitem[\protect\citeauthoryear{{Lau}, {Hearin}, {Nagai}  \&
  {Cappelluti}}{{Lau} et~al.}{2021}]{Lau_2021}
{Lau} E.~T.,  {Hearin} A.~P.,  {Nagai} D.,   {Cappelluti} N.,  2021, \mn@doi
  [\mnras] {10.1093/mnras/staa3313}, \href
  {https://ui.adsabs.harvard.edu/abs/2021MNRAS.500.1029L} {500, 1029}

\bibitem[\protect\citeauthoryear{{Leauthaud} et~al.,}{{Leauthaud}
  et~al.}{2016}]{leauthaud_etal:16}
{Leauthaud} A.,  et~al., 2016, \mn@doi [\mnras] {10.1093/mnras/stw117}, \href
  {http://adsabs.harvard.edu/abs/2016MNRAS.457.4021L} {457, 4021}

\bibitem[\protect\citeauthoryear{{Leauthaud} et~al.,}{{Leauthaud}
  et~al.}{2017}]{Leauthaud_2017}
{Leauthaud} A.,  et~al., 2017, \mn@doi [\mnras] {10.1093/mnras/stx258}, \href
  {https://ui.adsabs.harvard.edu/abs/2017MNRAS.467.3024L} {467, 3024}

\bibitem[\protect\citeauthoryear{{Lesgourgues} \& {Pastor}}{{Lesgourgues} \&
  {Pastor}}{2006}]{Lesgourgues_2006}
{Lesgourgues} J.,  {Pastor} S.,  2006, \mn@doi [\physrep]
  {10.1016/j.physrep.2006.04.001}, \href
  {https://ui.adsabs.harvard.edu/abs/2006PhR...429..307L} {429, 307}

\bibitem[\protect\citeauthoryear{{McClintock} et~al.,}{{McClintock}
  et~al.}{2019}]{McClintock_2018}
{McClintock} T.,  et~al., 2019, \mn@doi [\apj] {10.3847/1538-4357/aaf568},
  \href {http://adsabs.harvard.edu/abs/2019ApJ...872...53M} {872, 53}

\bibitem[\protect\citeauthoryear{{McEwen} \& {Weinberg}}{{McEwen} \&
  {Weinberg}}{2018}]{Mcewen_2018}
{McEwen} J.~E.,  {Weinberg} D.~H.,  2018, \mn@doi [\mnras]
  {10.1093/mnras/sty882}, \href
  {https://ui.adsabs.harvard.edu/abs/2018MNRAS.477.4348M} {477, 4348}

\bibitem[\protect\citeauthoryear{{Mezini}, {Zentner}, {Wang}  \&
  {Fielder}}{{Mezini} et~al.}{2024}]{Mezini_2024}
{Mezini} L.,  {Zentner} A.~R.,  {Wang} K.,   {Fielder} C.,  2024, \mn@doi
  [arXiv e-prints] {10.48550/arXiv.2406.10150}, \href
  {https://ui.adsabs.harvard.edu/abs/2024arXiv240610150M} {p. arXiv:2406.10150}

\bibitem[\protect\citeauthoryear{{Navarro}, {Frenk}  \& {White}}{{Navarro}
  et~al.}{1996}]{NFW_1996}
{Navarro} J.~F.,  {Frenk} C.~S.,   {White} S.~D.~M.,  1996, \mn@doi [\apj]
  {10.1086/177173}, \href {http://adsabs.harvard.edu/abs/1996ApJ...462..563N}
  {462, 563}

\bibitem[\protect\citeauthoryear{{Nishimichi} et~al.,}{{Nishimichi}
  et~al.}{2019}]{Nishimichi_2019}
{Nishimichi} T.,  et~al., 2019, \mn@doi [\apj] {10.3847/1538-4357/ab3719},
  \href {https://ui.adsabs.harvard.edu/abs/2019ApJ...884...29N} {884, 29}

\bibitem[\protect\citeauthoryear{{Padilla}, {Contreras}, {Zehavi}, {Baugh}  \&
  {Norberg}}{{Padilla} et~al.}{2019}]{Padilla_2019}
{Padilla} N.,  {Contreras} S.,  {Zehavi} I.,  {Baugh} C.~M.,   {Norberg} P.,
  2019, \mn@doi [\mnras] {10.1093/mnras/stz824}, \href
  {https://ui.adsabs.harvard.edu/abs/2019MNRAS.486..582P} {486, 582}

\bibitem[\protect\citeauthoryear{{Paillas} et~al.,}{{Paillas}
  et~al.}{2024}]{Paillas_2024}
{Paillas} E.,  et~al., 2024, \mn@doi [\mnras] {10.1093/mnras/stae1118}, \href
  {https://ui.adsabs.harvard.edu/abs/2024MNRAS.531..898P} {531, 898}

\bibitem[\protect\citeauthoryear{{Parejko} et~al.,}{{Parejko}
  et~al.}{2013}]{Parejko_LOWZ}
{Parejko} J.~K.,  et~al., 2013, \mn@doi [\mnras] {10.1093/mnras/sts314}, \href
  {http://adsabs.harvard.edu/abs/2013MNRAS.429...98P} {429, 98}

\bibitem[\protect\citeauthoryear{{Planck Collaboration} et~al.,}{{Planck
  Collaboration} et~al.}{2020}]{Planck_2020}
{Planck Collaboration} et~al., 2020, \mn@doi [\aap]
  {10.1051/0004-6361/201833910}, \href
  {https://ui.adsabs.harvard.edu/abs/2020A&A...641A...6P} {641, A6}

\bibitem[\protect\citeauthoryear{{Prada}, {Klypin}, {Cuesta}, {Betancort-Rijo}
  \& {Primack}}{{Prada} et~al.}{2012}]{Prada_2012}
{Prada} F.,  {Klypin} A.~A.,  {Cuesta} A.~J.,  {Betancort-Rijo} J.~E.,
  {Primack} J.,  2012, \mn@doi [\mnras] {10.1111/j.1365-2966.2012.21007.x},
  \href {https://ui.adsabs.harvard.edu/abs/2012MNRAS.423.3018P} {423, 3018}

\bibitem[\protect\citeauthoryear{{Reid}, {Seo}, {Leauthaud}, {Tinker}  \&
  {White}}{{Reid} et~al.}{2014}]{Reid_2014}
{Reid} B.~A.,  {Seo} H.-J.,  {Leauthaud} A.,  {Tinker} J.~L.,   {White} M.,
  2014, \mn@doi [\mnras] {10.1093/mnras/stu1391}, \href
  {http://adsabs.harvard.edu/abs/2014MNRAS.444..476R} {444, 476}

\bibitem[\protect\citeauthoryear{{Rogers}, {Peiris}, {Pontzen}, {Bird}, {Verde}
   \& {Font-Ribera}}{{Rogers} et~al.}{2019}]{Rogers_2019}
{Rogers} K.~K.,  {Peiris} H.~V.,  {Pontzen} A.,  {Bird} S.,  {Verde} L.,
  {Font-Ribera} A.,  2019, \mn@doi [\jcap] {10.1088/1475-7516/2019/02/031},
  \href {https://ui.adsabs.harvard.edu/abs/2019JCAP...02..031R} {2019, 031}

\bibitem[\protect\citeauthoryear{{Saito}, {Takada}  \& {Taruya}}{{Saito}
  et~al.}{2008}]{Saito_2008}
{Saito} S.,  {Takada} M.,   {Taruya} A.,  2008, \mn@doi [\prl]
  {10.1103/PhysRevLett.100.191301}, \href
  {https://ui.adsabs.harvard.edu/abs/2008PhRvL.100s1301S} {100, 191301}

\bibitem[\protect\citeauthoryear{{Salcedo}, {Maller}, {Berlind}, {Sinha},
  {McBride}, {Behroozi}, {Wechsler}  \& {Weinberg}}{{Salcedo}
  et~al.}{2018}]{Salcedo_2018}
{Salcedo} A.~N.,  {Maller} A.~H.,  {Berlind} A.~A.,  {Sinha} M.,  {McBride}
  C.~K.,  {Behroozi} P.~S.,  {Wechsler} R.~H.,   {Weinberg} D.~H.,  2018,
  \mn@doi [\mnras] {10.1093/mnras/sty109}, \href
  {https://ui.adsabs.harvard.edu/abs/2018MNRAS.475.4411S} {475, 4411}

\bibitem[\protect\citeauthoryear{{Salucci}}{{Salucci}}{2019}]{Salucci_2019}
{Salucci} P.,  2019, \mn@doi [\aapr] {10.1007/s00159-018-0113-1}, \href
  {https://ui.adsabs.harvard.edu/abs/2019A&ARv..27....2S} {27, 2}

\bibitem[\protect\citeauthoryear{{Skilling}}{{Skilling}}{2004}]{Skilling_2004}
{Skilling} J.,  2004, in {Fischer} R.,  {Preuss} R.,   {Toussaint} U.~V.,  eds,
   American Institute of Physics Conference Series Vol. 735, Bayesian Inference
  and Maximum Entropy Methods in Science and Engineering: 24th International
  Workshop on Bayesian Inference and Maximum Entropy Methods in Science and
  Engineering. pp 395--405, \mn@doi{10.1063/1.1835238}

\bibitem[\protect\citeauthoryear{{Smith}, {Watts}  \& {Sheth}}{{Smith}
  et~al.}{2006}]{Smith_2006}
{Smith} R.~E.,  {Watts} P.~I.~R.,   {Sheth} R.~K.,  2006, \mn@doi [\mnras]
  {10.1111/j.1365-2966.2005.09707.x}, \href
  {https://ui.adsabs.harvard.edu/abs/2006MNRAS.365..214S} {365, 214}

\bibitem[\protect\citeauthoryear{{Storey-Fisher}, {Tinker}, {Zhai}, {DeRose},
  {Wechsler}  \& {Banerjee}}{{Storey-Fisher} et~al.}{2022}]{Storey-Fisher_2022}
{Storey-Fisher} K.,  {Tinker} J.,  {Zhai} Z.,  {DeRose} J.,  {Wechsler} R.~H.,
   {Banerjee} A.,  2022, \mn@doi [arXiv e-prints] {10.48550/arXiv.2210.03203},
  \href {https://ui.adsabs.harvard.edu/abs/2022arXiv221003203S} {p.
  arXiv:2210.03203}

\bibitem[\protect\citeauthoryear{{Tinker} et~al.,}{{Tinker}
  et~al.}{2017}]{Tinker_2017a}
{Tinker} J.~L.,  et~al., 2017, \mn@doi [\apj] {10.3847/1538-4357/aa6845}, \href
  {https://ui.adsabs.harvard.edu/abs/2017ApJ...839..121T} {839, 121}

\bibitem[\protect\citeauthoryear{{Upadhye}, {Kwan}, {Pope}, {Heitmann},
  {Habib}, {Finkel}  \& {Frontiere}}{{Upadhye} et~al.}{2016}]{Upadhye_2016}
{Upadhye} A.,  {Kwan} J.,  {Pope} A.,  {Heitmann} K.,  {Habib} S.,  {Finkel}
  H.,   {Frontiere} N.,  2016, \mn@doi [\prd] {10.1103/PhysRevD.93.063515},
  \href {https://ui.adsabs.harvard.edu/abs/2016PhRvD..93f3515U} {93, 063515}

\bibitem[\protect\citeauthoryear{{Valogiannis} \& {Dvorkin}}{{Valogiannis} \&
  {Dvorkin}}{2022}]{Valogiannis_2022}
{Valogiannis} G.,  {Dvorkin} C.,  2022, \mn@doi [\prd]
  {10.1103/PhysRevD.106.103509}, \href
  {https://ui.adsabs.harvard.edu/abs/2022PhRvD.106j3509V} {106, 103509}

\bibitem[\protect\citeauthoryear{{Valogiannis}, {Yuan}  \&
  {Dvorkin}}{{Valogiannis} et~al.}{2024}]{Valogiannis_2024}
{Valogiannis} G.,  {Yuan} S.,   {Dvorkin} C.,  2024, \mn@doi [\prd]
  {10.1103/PhysRevD.109.103503}, \href
  {https://ui.adsabs.harvard.edu/abs/2024PhRvD.109j3503V} {109, 103503}

\bibitem[\protect\citeauthoryear{{Vega-Ferrero}, {Yepes}  \&
  {Gottl{\"o}ber}}{{Vega-Ferrero} et~al.}{2017}]{VegaFerrero_2017}
{Vega-Ferrero} J.,  {Yepes} G.,   {Gottl{\"o}ber} S.,  2017, \mn@doi [\mnras]
  {10.1093/mnras/stx282}, \href
  {https://ui.adsabs.harvard.edu/abs/2017MNRAS.467.3226V} {467, 3226}

\bibitem[\protect\citeauthoryear{{Viel}, {Haehnelt}  \& {Springel}}{{Viel}
  et~al.}{2010}]{Viel_2010}
{Viel} M.,  {Haehnelt} M.~G.,   {Springel} V.,  2010, \mn@doi [\jcap]
  {10.1088/1475-7516/2010/06/015}, \href
  {https://ui.adsabs.harvard.edu/abs/2010JCAP...06..015V} {2010, 015}

\bibitem[\protect\citeauthoryear{{Wechsler} \& {Tinker}}{{Wechsler} \&
  {Tinker}}{2018}]{Wechsler_2018}
{Wechsler} R.~H.,  {Tinker} J.~L.,  2018, \mn@doi [\araa]
  {10.1146/annurev-astro-081817-051756}, \href
  {https://ui.adsabs.harvard.edu/abs/2018ARA&A..56..435W} {56, 435}

\bibitem[\protect\citeauthoryear{{White}, {Zheng}, {Brown}, {Dey}  \&
  {Jannuzi}}{{White} et~al.}{2007}]{White_2007}
{White} M.,  {Zheng} Z.,  {Brown} M. J.~I.,  {Dey} A.,   {Jannuzi} B.~T.,
  2007, \mn@doi [\apjl] {10.1086/512015}, \href
  {https://ui.adsabs.harvard.edu/abs/2007ApJ...655L..69W} {655, L69}

\bibitem[\protect\citeauthoryear{{White} et~al.,}{{White}
  et~al.}{2011}]{CMASS_Martin}
{White} M.,  et~al., 2011, \mn@doi [\apj] {10.1088/0004-637X/728/2/126}, \href
  {http://adsabs.harvard.edu/abs/2011ApJ...728..126W} {728, 126}

\bibitem[\protect\citeauthoryear{{Wibking} et~al.,}{{Wibking}
  et~al.}{2017}]{Wibking_2017}
{Wibking} B.~D.,  et~al., 2017, preprint, \href
  {http://adsabs.harvard.edu/abs/2017arXiv170907099W} {} (\mn@eprint {arXiv}
  {1709.07099})

\bibitem[\protect\citeauthoryear{{Wibking}, {Weinberg}, {Salcedo}, {Wu},
  {Singh}, {Rodr{\'\i}guez-Torres}, {Garrison}  \& {Eisenstein}}{{Wibking}
  et~al.}{2020}]{Wibking_2020}
{Wibking} B.~D.,  {Weinberg} D.~H.,  {Salcedo} A.~N.,  {Wu} H.-Y.,  {Singh} S.,
   {Rodr{\'\i}guez-Torres} S.,  {Garrison} L.~H.,   {Eisenstein} D.~J.,  2020,
  \mn@doi [\mnras] {10.1093/mnras/stz3423}, \href
  {https://ui.adsabs.harvard.edu/abs/2020MNRAS.492.2872W} {492, 2872}

\bibitem[\protect\citeauthoryear{{Xu}, {Zehavi}  \& {Contreras}}{{Xu}
  et~al.}{2020}]{Xu_2020}
{Xu} X.,  {Zehavi} I.,   {Contreras} S.,  2020, arXiv e-prints, \href
  {https://ui.adsabs.harvard.edu/abs/2020arXiv200705545X} {p. arXiv:2007.05545}

\bibitem[\protect\citeauthoryear{{Yang}, {Mo}, {van den Bosch}, {Jing},
  {Weinmann}  \& {Meneghetti}}{{Yang} et~al.}{2006}]{Yang_2006}
{Yang} X.,  {Mo} H.~J.,  {van den Bosch} F.~C.,  {Jing} Y.~P.,  {Weinmann}
  S.~M.,   {Meneghetti} M.,  2006, \mn@doi [\mnras]
  {10.1111/j.1365-2966.2006.11091.x}, \href
  {https://ui.adsabs.harvard.edu/abs/2006MNRAS.373.1159Y} {373, 1159}

\bibitem[\protect\citeauthoryear{{York} et~al.,}{{York}
  et~al.}{2000}]{SDSS_York}
{York} D.~G.,  et~al., 2000, \mn@doi [\aj] {10.1086/301513}, \href
  {http://adsabs.harvard.edu/abs/2000AJ....120.1579Y} {120, 1579}

\bibitem[\protect\citeauthoryear{{Yuan}, {Hadzhiyska}, {Bose}, {Eisenstein}  \&
  {Guo}}{{Yuan} et~al.}{2020}]{Yuan_2020}
{Yuan} S.,  {Hadzhiyska} B.,  {Bose} S.,  {Eisenstein} D.~J.,   {Guo} H.,
  2020, arXiv e-prints, \href
  {https://ui.adsabs.harvard.edu/abs/2020arXiv201004182Y} {p. arXiv:2010.04182}

\bibitem[\protect\citeauthoryear{{Yuan}, {Garrison}, {Eisenstein}  \&
  {Wechsler}}{{Yuan} et~al.}{2022}]{Yuan_2022}
{Yuan} S.,  {Garrison} L.~H.,  {Eisenstein} D.~J.,   {Wechsler} R.~H.,  2022,
  \mn@doi [\mnras] {10.1093/mnras/stac1830}, \href
  {https://ui.adsabs.harvard.edu/abs/2022MNRAS.515..871Y} {515, 871}

\bibitem[\protect\citeauthoryear{{Yuan}, {Abel}  \& {Wechsler}}{{Yuan}
  et~al.}{2024}]{Yuan_2024}
{Yuan} S.,  {Abel} T.,   {Wechsler} R.~H.,  2024, \mn@doi [\mnras]
  {10.1093/mnras/stad3359}, \href
  {https://ui.adsabs.harvard.edu/abs/2024MNRAS.527.1993Y} {527, 1993}

\bibitem[\protect\citeauthoryear{{Zel'dovich}}{{Zel'dovich}}{1970}]{Zeldovich_1970}
{Zel'dovich} Y.~B.,  1970, \aap, \href
  {https://ui.adsabs.harvard.edu/abs/1970A&A.....5...84Z} {5, 84}

\bibitem[\protect\citeauthoryear{{Zemp}, {Gnedin}, {Gnedin}  \&
  {Kravtsov}}{{Zemp} et~al.}{2011}]{Zemp_2011}
{Zemp} M.,  {Gnedin} O.~Y.,  {Gnedin} N.~Y.,   {Kravtsov} A.~V.,  2011, \mn@doi
  [\apjs] {10.1088/0067-0049/197/2/30}, \href
  {https://ui.adsabs.harvard.edu/abs/2011ApJS..197...30Z} {197, 30}

\bibitem[\protect\citeauthoryear{{Zhai} et~al.,}{{Zhai}
  et~al.}{2017a}]{Zhai_2017}
{Zhai} Z.,  et~al., 2017a, \mn@doi [\apj] {10.3847/1538-4357/aa8eee}, \href
  {http://adsabs.harvard.edu/abs/2017ApJ...848...76Z} {848, 76}

\bibitem[\protect\citeauthoryear{{Zhai}, {Blanton}, {Slosar}  \&
  {Tinker}}{{Zhai} et~al.}{2017b}]{Zhai_2017c}
{Zhai} Z.,  {Blanton} M.,  {Slosar} A.,   {Tinker} J.,  2017b, \mn@doi [\apj]
  {10.3847/1538-4357/aa9888}, \href
  {https://ui.adsabs.harvard.edu/abs/2017ApJ...850..183Z} {850, 183}

\bibitem[\protect\citeauthoryear{{Zhai} et~al.,}{{Zhai}
  et~al.}{2019}]{Zhai_2019}
{Zhai} Z.,  et~al., 2019, \mn@doi [\apj] {10.3847/1538-4357/ab0d7b}, \href
  {https://ui.adsabs.harvard.edu/abs/2019ApJ...874...95Z} {874, 95}

\bibitem[\protect\citeauthoryear{{Zhai}, {Percival}  \& {Guo}}{{Zhai}
  et~al.}{2023a}]{Zhai_2023b}
{Zhai} Z.,  {Percival} W.~J.,   {Guo} H.,  2023a, \mn@doi [\mnras]
  {10.1093/mnras/stad1793}, \href
  {https://ui.adsabs.harvard.edu/abs/2023MNRAS.523.5538Z} {523, 5538}

\bibitem[\protect\citeauthoryear{{Zhai} et~al.,}{{Zhai}
  et~al.}{2023b}]{Zhai_2023a}
{Zhai} Z.,  et~al., 2023b, \mn@doi [\apj] {10.3847/1538-4357/acc65b}, \href
  {https://ui.adsabs.harvard.edu/abs/2023ApJ...948...99Z} {948, 99}

\bibitem[\protect\citeauthoryear{{Zheng} et~al.,}{{Zheng}
  et~al.}{2005}]{Zheng_2005}
{Zheng} Z.,  et~al., 2005, \mn@doi [\apj] {10.1086/466510}, \href
  {http://adsabs.harvard.edu/abs/2005ApJ...633..791Z} {633, 791}

\bibitem[\protect\citeauthoryear{{Zheng}, {Coil}  \& {Zehavi}}{{Zheng}
  et~al.}{2007}]{Zheng_2007}
{Zheng} Z.,  {Coil} A.~L.,   {Zehavi} I.,  2007, \mn@doi [\apj]
  {10.1086/521074}, \href {http://adsabs.harvard.edu/abs/2007ApJ...667..760Z}
  {667, 760}

\bibitem[\protect\citeauthoryear{{Zu}, {Zheng}, {Zhu}  \& {Jing}}{{Zu}
  et~al.}{2008}]{Zu_2008}
{Zu} Y.,  {Zheng} Z.,  {Zhu} G.,   {Jing} Y.~P.,  2008, \mn@doi [\apj]
  {10.1086/591071}, \href
  {https://ui.adsabs.harvard.edu/abs/2008ApJ...686...41Z} {686, 41}

\bibitem[\protect\citeauthoryear{{van Daalen}, {Angulo}  \& {White}}{{van
  Daalen} et~al.}{2012}]{vanDaalen_2012}
{van Daalen} M.~P.,  {Angulo} R.~E.,   {White} S. D.~M.,  2012, \mn@doi
  [\mnras] {10.1111/j.1365-2966.2012.21437.x}, \href
  {https://ui.adsabs.harvard.edu/abs/2012MNRAS.424.2954V} {424, 2954}

\bibitem[\protect\citeauthoryear{van~der Walt, Colbert  \& Varoquaux}{van~der
  Walt et~al.}{2011}]{numpy}
van~der Walt S.,  Colbert S.~C.,   Varoquaux G.,  2011, \mn@doi [Computing in
  Science Engineering] {10.1109/MCSE.2011.37}, 13, 22

\makeatother
\end{thebibliography}

\bsp	% typesetting comment
\label{lastpage}
\end{document}